\documentclass[iop,apj]{emulateapj}

\shorttitle{Timing and Scattering of PALFA Pulsars}
\shortauthors{Nice et al.}

\submitted{Astrophysical Journal, 772: 50, (Received 2013 April 26; accepted 2013 June 3; published 2013 July 5)}

\begin{document}

\title{Timing and Interstellar Scattering of 35 Distant Pulsars Discovered in the PALFA Survey}

\author{
D.~J.~Nice\altaffilmark{1},
E.~Altiere\altaffilmark{2,3},
S.~Bogdanov\altaffilmark{4},
J.~M.~Cordes\altaffilmark{5},
D.~Farrington\altaffilmark{2},
J.~W.~T.~Hessels\altaffilmark{6,7},
V.~M.~Kaspi\altaffilmark{8},
A.~G.~Lyne\altaffilmark{9}, 
L.~Popa\altaffilmark{2,10},
S.~M.~Ransom\altaffilmark{11},
S.~Sanpa-arsa\altaffilmark{12}
B.~W.~Stappers\altaffilmark{9},
Y.~Wang\altaffilmark{2},
B.~Allen\altaffilmark{13,14,15},
N.~D.~R.~Bhat\altaffilmark{16}, 
A.~Brazier\altaffilmark{5},
F.~Camilo\altaffilmark{4,17},
D.~J.~Champion\altaffilmark{18},
S.~Chatterjee\altaffilmark{5},
F.~Crawford\altaffilmark{19},
J.~S.~Deneva\altaffilmark{17},
G.~Desvignes\altaffilmark{20},
P.~C.~C.~Freire\altaffilmark{18},
F.~A.~Jenet\altaffilmark{21}, 
B.~Knispel\altaffilmark{13,14},  
P.~Lazarus\altaffilmark{18}, 
K.~J.~Lee\altaffilmark{18},
J.~van~Leeuwen\altaffilmark{6,7}, 
D.~R.~Lorimer\altaffilmark{22},
R.~Lynch\altaffilmark{8}, 
M.~A.~McLaughlin\altaffilmark{22},
P.~Scholz\altaffilmark{8},
X.~Siemens\altaffilmark{15}, 
I.~H.~Stairs\altaffilmark{3}, 
K.~Stovall\altaffilmark{21}, 
A.~Venkataraman\altaffilmark{17},
W.~Zhu\altaffilmark{3}
}

\altaffiltext{1}{Dept.~of Physics, Lafayette College, Easton, PA 18042
, USA
} 
\altaffiltext{2}{Dept.~of Physics, Bryn Mawr College, Bryn Mawr, PA 19010, USA}
\altaffiltext{3}{Dept.~of Physics and Astronomy, Univ.\ of British Columbia, Vancouver, BC V6T 1Z1, Canada} 
\altaffiltext{4}{Columbia Astrophysics Laboratory, Columbia Univ.\,  New York, NY 10027, USA} 
\altaffiltext{5}{Astronomy Dept.,  Cornell Univ.\, Ithaca, NY 14853, USA} 
\altaffiltext{6}{ASTRON, The Netherlands Institute for Radio Astronomy, Postbus 2, 7990 AA, Dwingeloo, The Netherlands} 
\altaffiltext{7}{Astronomical Institute ``Anton Pannekoek'', Univ.\ of Amsterdam, Science Park 904, 1098 XH Amsterdam, The Netherlands}
\altaffiltext{8}{Dept.~of Physics, McGill Univ.\, Montreal, QC H3A 2T8, Canada}
\altaffiltext{9}{Jodrell Bank Centre for Astrophysics, School of Physics and Astronomy, Univ.\ of Manchester, Manchester, M13 9PL, UK} 
\altaffiltext{10}{Dept.~of Physics, Massachusetts Institute of Technology, Cambridge, MA 02139 USA}
\altaffiltext{11}{NRAO, Charlottesville, VA 22903, USA} 
\altaffiltext{12}{Dept.~of Astronomy, Univ.\ of Virginia, Charlottesville, VA, 22903, USA}
\altaffiltext{13}{Max-Planck-Institut f\"ur Gravitationsphysik, D-30167 Hannover, Germany}
\altaffiltext{14}{Leibniz Universit{\"a}t Hannover, D-30167 Hannover, Germany}
\altaffiltext{15}{Physics Dept.~ Univ.\ of Wisconsin -- Milwaukee, Milwaukee WI 53211, USA}
\altaffiltext{16}{Center for Astrophysics and Supercomputing, Swinburne Univ.\, Hawthorn, Victoria 3122, Australia} 
\altaffiltext{17}{Arecibo Observatory, HC3 Box 53995, Arecibo, PR 00612
, USA
}
\altaffiltext{18}{Max-Planck-Institut f\"ur Radioastronomie, D-53121 Bonn, Germany} 
\altaffiltext{19}{Dept.~of Physics and Astronomy, Franklin and Marshall College, Lancaster, PA 17604-3003, USA} 
\altaffiltext{20}{Dept.~of Astronomy and Radio Astronomy Laboratory, Univ.\ of California, Berkeley, CA 94720, USA} 
\altaffiltext{21}{Center for Gravitational Wave Astronomy, Univ.\ of Texas at Brownsville, TX 78520, USA} 
\altaffiltext{22}{Dept.~of Physics, West Virginia Univ.\, Morgantown, WV 26506, USA} 

\begin{abstract}

We have made extensive observations of 35 distant slow (non-recycled) pulsars discovered 
in the ongoing Arecibo PALFA pulsar survey.  Timing observations of these pulsars
over several years at Arecibo Observatory and Jodrell Bank Observatory have yielded
high-precision positions and measurements of rotation properties.
Despite being a relatively distant population, 
these pulsars have properties that 
mirror those of the previously known pulsar population.
Many of the sources exhibit timing noise, and one underwent a small glitch.
We have used multifrequency data to measure
the interstellar scattering properties of these pulsars.
We find scattering to be higher than predicted along some
lines of sight, particularly in
the Cygnus region.
Lastly, we 
present {\it XMM-Newton} and {\it Chandra} observations
of the youngest and most energetic of the pulsars, J1856+0245, which has previously been
associated with the GeV--TeV pulsar wind nebula HESS
J1857+026.

\end{abstract}


\section{Introduction}\label{sec:intro}

More than 2000 pulsars are now known \citep{hm13}.  Most of these
were discovered in blind surveys by radio telescopes.  In these surveys,
short individual
telescope pointings, of order a few minutes, are made over grids of
thousands of sky positions.  In each pointing of a typical survey, 
the radio frequency signal is divided into hundreds of narrow spectral 
channels, 
each of which is 
square-law detected and
sampled at a rate of order 10~kHz.  The resulting
data blocks are searched for dispersed signals exhibiting either 
periodic or transient behavior expected of pulsars.   For any newly
discovered pulsar, survey data immediately yield approximate estimates of the 
pulsar's position (with the precision of the telescope primary beam size,
typically between a few arc minutes and a degree), along with the pulsar's
period and dispersion measure (DM).

Follow-up timing observations of newly discovered pulsars
are needed to characterize more precisely the pulsar and its environment.  
Observations spaced over several months or longer decouple 
the influences of pulsar 
rotation and time of flight across the solar system,
allowing measurement of pulsar spin-down rates
and positions, the latter with
sub-arcsecond precision.
Such timing observations can
also detect orbital motion of pulsars in binary systems.
Measured pulsar spin-down rates can be used
to estimate the pulsar's age, magnetic field strength,
and energy loss rate.  
Measured pulsar positions can be used
to search for counterparts to the pulsar in other spectral bands,
and phase coherent timing solutions can 
aid searches for pulsed radiation there.

The Pulsar Arecibo L-Band Feed Array (PALFA) project is an ongoing 
deep pulsar survey of low Galactic latitudes being undertaken at
the 305-m William E.\ Gordon Telescope at the Arecibo
Observatory  
\citep{cfl+06,lab+12}.
Among the survey goals are 
discovering 
millisecond pulsars suitable
for use in pulsar timing arrays that will enable the detection of
gravitational waves,
discovering binary pulsars useful for tests of strong-field gravitation
and determination of the neutron star equation of state,
and discovering distant pulsars
to better characterize the pulsar population 
and the interstellar medium on 
Galactic scales.

In this paper, we describe several years
of follow-up timing observations of 35 slow (non-recycled) pulsars 
discovered in the PALFA survey between
2004 and 2008.  These are the first published observations
of most of these sources, although a few of the discoveries
have been announced previously either as conventional
pulsars \citep{cfl+06,hng+08} or rotating radio
transients \citep{dcm+09}.
The timing observations were made at
both the Arecibo Observatory and the Jodrell Bank Observatory.
We give a synopsis of the PALFA survey
in Section~\ref{sec:survey}.  We describe the 
observations in Section~\ref{sec:obs}.  We present
timing models of the pulsars in Section~\ref{sec:timing}.
We give
flux densities and spectral indices of the pulsars 
measured from the timing data in Section~\ref{sec:flux}.
We describe the pulse profiles and use them
to characterize the scattering
of the pulsar signals by the interstellar medium in
Section~\ref{sec:profilesandscattering}.  
We describe dedicated X-ray observations of the young pulsar
J1856+0245 in Section~\ref{sec:1856xray}.  We discuss our results in
Section~\ref{sec:discussion}.

\section{The PALFA survey}\label{sec:survey}

The parameters of the PALFA survey are discussed in
\citet{cfl+06}, \citet{vcl+06}, and \citet{lab+12}.
A detailed description of the early survey observations will shortly
be published  
(J.~Swiggum et al.~2013, in preparation).
We give a brief summary here.

The survey area covers those portions of low Galactic
latitude sky, 
$|b|<5^{\circ}$, visible with the Arecibo telescope.
This includes Galactic
longitude ranges
$32^{\circ}  \lesssim \ell \lesssim  77 ^{\circ}$ and 
$168{^\circ} \lesssim \ell \lesssim 214^{\circ}$.
Survey observations used the ALFA (Arecibo L-Band Feed Array) 
receiver, which provides seven 
simultaneous independent beams on the sky.
Data were collected over 100\,MHz passbands centered
on 1420\,MHz in each receiver beam.  Each passband
was processed by a
Wideband Arecibo Pulsar Processor (WAPP) three-level 
autocorrelator system (see also Section~\ref{sec:obs}).  
Autocorrelation spectra with 256 lags were recorded at
intervals of 64~$\mu$s, with orthogonal polarizations
summed before recording.  The typical dwell time for
a telescope pointing was 268\,s.  (More recent PALFA
survey observations, made after the discoveries
discussed in this paper, have used upgraded spectrometers, providing
960 channels across 320\,MHz passbands.)

Previous surveys of the low Galactic latitude sky
visible from Arecibo \citep{ht75b,sstd86,nft95}
used lower radio frequencies and were constrained
by the narrow bandwidths available on the telescope before
its upgrade in the mid-1990s.  In contrast, the PALFA
survey's relatively high radio frequency of 1420~MHz, 
relatively large bandwidth, and fine spectral resolution
give it sensitivity to pulsars over
a far larger volume of the Galaxy than ever before,
particularly to those pulsars with
short spin periods and high DMs \citep{csl+12}.

The bulk of the pulsars observed for this paper were discovered by the
``quick look'' pipeline, a low-resolution periodicity search,
which ran in near-real-time as data were collected \citep{cfl+06}.  The 
remaining pulsars 
was discovered either by full-resolution periodicity searches
or by searches for dispersed transient signals \citep{cfl+06,dcm+09}.

While the present paper focuses on isolated long-period pulsars, the PALFA
survey has also found numerous binary and millisecond pulsars
\citep[e.g.,][]{lsf+06, crl+08, kac+10, kla+11, dfc+12, csl+12, akc+13}.

\section{Timing Observations}\label{sec:obs}

We observed all 35 pulsars in this paper at the
Arecibo Observatory.  In addition, we observed 17 of the 
stronger sources at Jodrell Bank Observatory.  
Details of observations for individual sources 
(observatory used, cadence, time span of observation) can be inferred
from plots presented in Section~\ref{sec:timing}.
Although several of the pulsars in this paper were discovered
through detection of individual pulses, all timing for this paper
was done by averaging pulsar time series over many pulse periods.

Observations at Arecibo Observatory
used the ``L-wide'' receiver, with system equivalent flux
density of approximately 2.4--3.0~Jy, depending on zenith angle, 
frequency, and sky
temperature.  Data were collected using four WAPPs 
three-level autocorrelation spectrometers \citep{dsh00}.   
The spectrometers operated in parallel to cover four receiver subbands,
centered at frequencies of 1170, 1370, 1470, and 
1570~MHz, with 100~MHz bandwidth in each subband.
Within each subband, 256-lag autocorrelations were accumulated and recorded
at intervals of 256~$\mu$s.  Self- and cross-products of two circular orthogonal
polarizations were recorded, although only the self-products were used
for the present work.  A typical observation duration was 2--3 minutes.  
A pulsed radio-frequency reference signal of known strength was injected into
the receiver and recorded prior to the observation of each pulsar 
in order to calibrate the receiver amplification, as needed to balance the
two orthogonal polarizations and to measure absolute flux
densities.  Absolute flux calibration used measurements of the reference
signal strength provided by the Observatory.

The recorded spectral time series were dedispersed and folded off-line using
a modified version of the {\sc Sigproc} software 
package\footnote{\url{http://sigproc.sourceforge.net}}.
Data from each subband were processed separately.   Observations were
processed in subintervals of 30~s, 
so a typical 2-3~minute observation in four subbands 
yielded 16-24 folded pulse profiles.  Each of these profiles
was cross-correlated with a reference profile to produce a time of arrival
(TOA) using standard techniques \citep[e.g.,][]{lk05}.  
Thus, the Arecibo data set for any
given observation of a source consisted of order two dozen TOAs.
The same reference profile was used to process data from all four subbands, 
even for pulsars with significant profile evolution.
The TOA sets for some weak pulsars included a significant number of
TOAs which did not fit timing models (described below) because of
signal-to-noise ratios too low to robustly detect the pulsar signal.  
The data sets were edited using manual and semi-automated
methods to remove obviously spurious TOAs.

Observations at Arecibo Observatory were made at roughly six-week intervals
between 2006 March and 2010 March except for a gap between 2007 March 
and 2007 December due to
extended telescope maintenance.  A few additional 
observations were made in 2012 September.  Within an observing session, 
any given pulsar was observed either on a single day or on a pair of 
adjacent days.  Occasionally pulsars were observed multiple times within
a day to resolve 
pulse numbering 
ambiguities.

Observations at Jodrell Bank used a dual-polarization cryogenic
receiver on the 76-m Lovell telescope, with a system equivalent flux
density of 25 Jy on cold sky.  Data were processed by a digital filterbank
which covered the frequency band between 1350 MHz and 1700 MHz with
channels of 0.5 MHz bandwidth. Observations were typically made with a
total duration of between 10 minutes and 40 minutes, depending upon the
discovery signal-to-noise ratio. Data were folded at the nominal
topocentric period of the pulsar for subintegration times of 10~s.
After inspection and ``cleaning'' of any radio-frequency interference, the
profiles were dedispersed at the nominal value of the pulsar DM.
Initial
pulsar parameters were established by conducting local searches in period
and DM about the nominal discovery values and 
finally summed over frequency
and time to produce integrated profiles.  TOAs were obtained after matching
with a standard template and processed using standard analysis techniques
with {\sc Psrtime}\footnote{\url{http://www.jb.man.ac.uk/pulsar/observing/progs/psrtime.html}} and {\sc Tempo}\footnote{\url{http://tempo.sourceforge.net}}.

Observations at Jodrell Bank were made between 2005 June and 2012 April. Observations
of a given source were made at intervals of 1-2 weeks.  The observation
of any given source yielded a single TOA
on a given day.

\section{Timing}\label{sec:timing}

\begin{figure*}[p]
\includegraphics[width=0.80\linewidth]{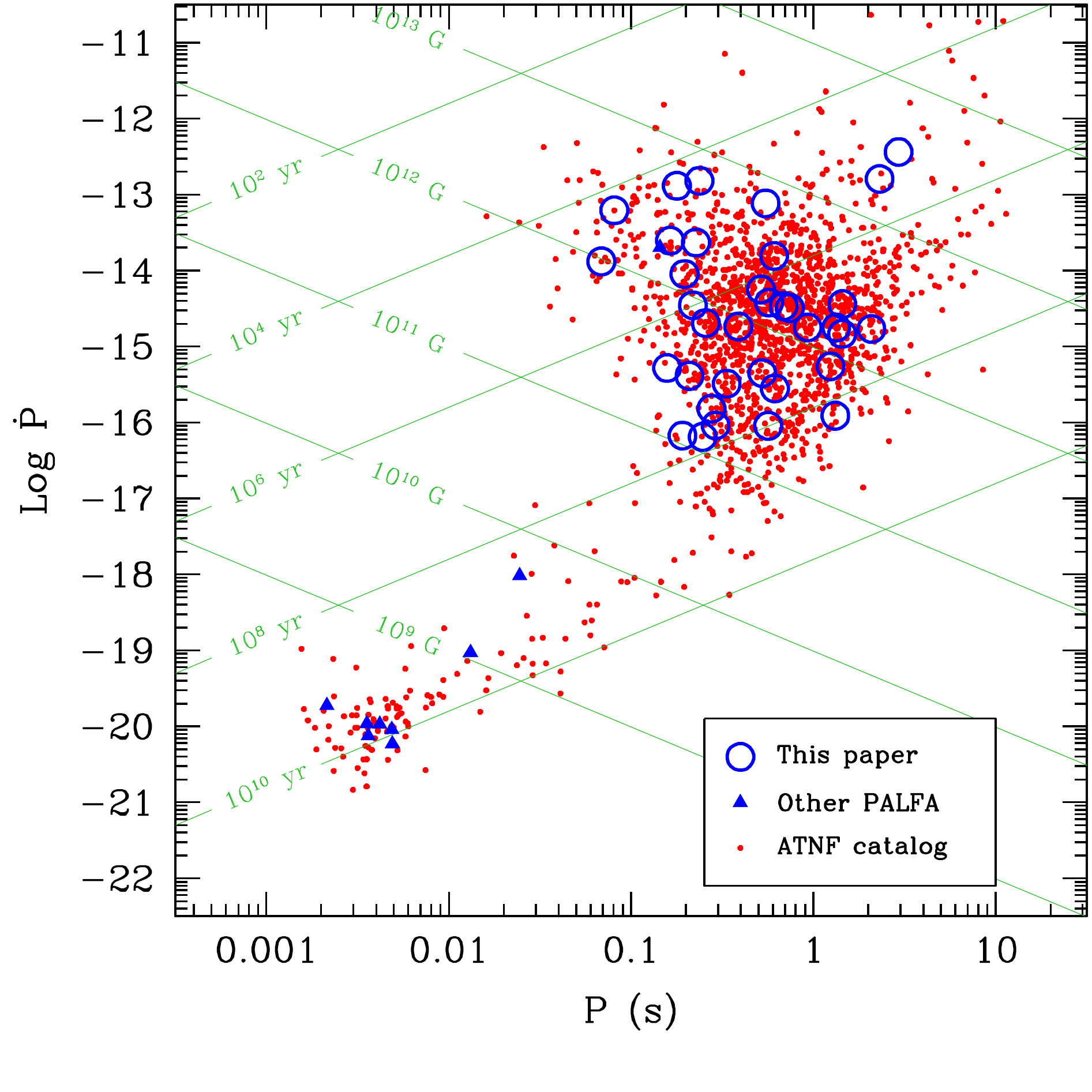}
\caption{\label{fig:ppdot}Pulsar rotation periods and period derivatives.
Blue circles 
indicate pulsars in the present paper, 
blue triangles 
indicate other published pulsars discovered in the PALFA survey, and 
red dots 
are all known pulsars in the Galactic disk
\citep{hm13}.
}
\end{figure*}

The TOAs from Arecibo and Jodrell Bank (where applicable) were combined into
a single data set for each pulsar, which was fit to a standard timing model
using $\chi^2$ minimization techniques implemented in
the {\sc Tempo} software
package.  The basic timing model for each source was parameterized
by right ascension,
$\alpha$; declination, $\delta$; pulse phase; rotation period, $P$;
rotation period time derivative, $\dot{P}$; and dispersion measure, DM.
Additional parameters were fit to model timing noise in some pulsars,
and a glitch in one pulsar, as described below.
The best-fit parameters for each pulsar are listed in Table~\ref{table:parameters}.
Parameter uncertainties listed in the table are twice the formal uncertainties
calculated in the timing model fit.  Spin periods and period derivatives
are plotted in Figure~\ref{fig:ppdot}.  Values of DM given in this
table do not include corrections for scattering or intrinsic profile evolution
(Section~\ref{sec:scattering}).

For pulsars observed at Jodrell Bank, an arbitrary time offset was allowed
between the Arecibo and Jodrell Bank TOAs.  The timing analysis used the
JPL DE405 solar system ephemeris \citep{sta98b} and the TT(BIPM11) time
scale\footnote{\url{ftp://tai.bipm.org/TFG/TT(BIPM)/}}.  Because of
the prevalence of timing noise (see below), unweighted fits were used.

Residual pulse arrival times from fits of each pulsar to 
the basic timing model are shown in
Figures~\ref{fig:resid1}(a)-(e).
Timing noise is evident in many sources.  To measure
accurate pulsar parameters, TOAs for these sources 
were ``whitened'' by including Taylor expansionss of the pulsar
periods in the timing models, $P(t)=\sum_{n=0}^{n_{\rm fit}} (1/n!)P^{(n)}(t-t_0)^n$,
where $t$ is the time, $t_0$ is the period epoch, and $P^{(n)}\equiv [d^nP/dt^n]_{t_0}$.
For each pulsar, models with $n_{\rm fit}$ from 1 (the basic timing model) through 10 
were tried, and an appropriate value of $n_{\rm fit}$ was chosen
based on the standard deviations of the residuals and on visual inspection
of the residuals to ensure there were not groups of correlated nonzero residuals.
The chosen values of $n_{\rm fit}$
for each pulsar are listed in Table~\ref{table:parameters}.

For four pulsars, designated as ``N'' in the $n_{\rm fit}$ 
column of Table~\ref{table:parameters}, polynomial models with $n_{\rm fit}$ as high as 10 
did not yield white residuals.  An alternative method was used to estimate
the value of each timing parameter for these pulsars.  For each pulsar, the timing
model (including 10th-order polynomial in $P(t))$ which minimized $\chi^2$ was found.  
Each parameter was then individually perturbed (allowing other parameters to freely vary
in the fit) to find the parameter values which resulted in a threefold increase in 
the standard deviation of the residuals.  The resulting range of parameter values was
used to set the uncertainties presented in Table~\ref{table:parameters}.
We view this as a highly conservative approach to estimating the uncertainties.

Because many of the pulse periods were analyzed as high-order Taylor series around the 
period epochs, the values of $P$ and $\dot{P}$ listed in Table~\ref{table:parameters}
are the best estimates of the values at the period epoch; they are not
averages over the length of the data set.

In Table~\ref{table:derived} we present several quantities derived from the 
timing model parameters.  
Estimates of spin-down age, $t_{\rm s}$, surface magnetic
field strength, $B$, and spin-down energy loss rates, $\dot{E}$, were 
made using conventional formulae \citep{lk05} and are given in
logarithmic form in the table.
Positions in Galactic coordinates are based on the
positions in equatorial coordinates in Table~\ref{table:scatter}.  
Distances are estimated from dispersion measured in Table~\ref{table:parameters}
using the NE2001 model of the distribution of ionized material in the
Galaxy \citep{cl02}.  The positions of the pulsars projected onto the Galactic
plane are shown in Figure~\ref{fig:galplot}.

We detected one glitch, in PSR~J1947+1957, as evinced by the peak in the residual
pulse arrival times for this pulsar in Figure~\ref{fig:resid1}(d).  The glitch event
was on MJD~$55085.5\pm2.6$ and was well fit by 
an instantaneous, permanent change in rotation frequency of 
$\Delta \nu=(1.02\pm0.03)\times 10^{-8}\,{\rm s}^{-1}$, which is a fractional
change of $\Delta \nu/\nu=(1.61\pm0.04)\times 10^{-9}$.  This is a typical
value for a small pulsar glitch \citep[e.g.,][]{elsk11}.  Other pulsars in our data set
may have had glitch-like events; in particular, see the residual arrival times
of PSRs~J1905+0902 and~J1916+12245 in Figure~\ref{fig:resid1}(b).  However, due to
the small sizes of the events and the coarse sampling of the time series, 
these events cannot easily be distinguished from smoothly-varying timing noise.

\section{Fluxes and spectral indices}\label{sec:flux}

Flux densities and spectral indices of the pulsars are given in Table~\ref{table:flux}.
These were measured from the calibrated
pulse profiles from the Arecibo observations by the following procedure.
(For logistical
reasons, only data from some observing dates were used.)  Each profile
was generated from a 30~s observation in one of the four 100~MHz subbands,
as described above.  The flux density was measured in each profile.
All flux density measurements for
a given subband in a given day were combined; the resulting daily average
values were then combined to find a global average for that subband.
Finally, the four average flux densities from the four subbands were fit
to a function of the form $S(f)=S_{1400}(f_{\rm MHz}/1400)^{\alpha}$, with flux 
density $S_{1400}$ and spectral index $\alpha$ as free parameters in
the fit.  The best-fit parameters for each pulsar are given in Table~\ref{table:flux},
along with the reduced chi-squared values, $\chi^2_\nu$, of the fits.
Although the formal uncertainties from the fits are reported in the table, 
they should be interpreted with caution, both because they arise from
fits with only two  degrees of freedom (four data points and two fit parameters)
and because scintillation could cause 
apparent, non-intrinsic
variations in pulsar flux 
densities from epoch to epoch.  Table~\ref{table:flux} also includes luminosity estimates,
$L_{1400}=S_{1400}d^2$, using the distances derived from DMs
as reported in Table~\ref{table:scatter}.

\section{Profile morphologies and scattering properties}\label{sec:profilesandscattering}

\subsection{Profiles}\label{sec:profiles}

Profiles of each pulsar in each of the four observing subbands are given in Figures~\ref{fig:profilea}--\ref{fig:profilecloseupa}.
These average profiles were generated by coherently
adding all Arecibo data profiles for a given pulsar that yielded good TOAs.  The alignment of the profiles was made
using the timing solutions described in Section~\ref{sec:timing}.
By and large, these profiles are typical of those seen in the previously known pulsar population.
Some evidence of profile evolution over frequency is seen (e.g., relative
component strengths of PSR J1948+2551).

One source, J1909+0749, has a strong interpulse.  Of its two pulse components,
we have somewhat arbitrarily designated the shorter, broader component
(near the center of its profiles in Figure~\ref{fig:profilea}(a)) as the main pulse
and the taller, narrower component (near the end of its profiles in Figure~\ref{fig:profilea}(a))
as the interpulse.  
The interpulse trails the main pulse by $\sim 0.484$ of a pulse period (174$^\circ$).  
The ratio of integrated flux density for the main pulse over that
of the interpulse in each of the Arecibo observing subbands is 1.28, 1.21, 1.02, and 1.36, 
at 1170, 1370, 1470, and 1570~MHz, respectively.  We do not know of a physical explanation 
for the dip in this ratio at 1470~MHz.

The profiles in Figures~\ref{fig:profilea}(a) and (b) were used to compute 
pulse widths at half-maximum, $W_{50}$, for each pulsar in each observing band.
The values are given in Table~\ref{table:width}, both in milliseconds and as
a fraction of a pulse period.

\subsection{Scattering}\label{sec:scattering}

The profiles of several pulsars show evidence of frequency-dependent pulse broadening,
indicative of multipath scattering in the interstellar medium.   The observed profiles
result from the convolution of pulse-broadening functions (PBFs) with
the intrinsic pulse shapes.
To quantify the scattering for each pulsar's signal, we used least-squares techniques to
simultaneously fit PBFs to the profiles of the pulsar in each of the four
observing subbands. 
While PBFs can take
on a range of shapes dependent on the distribution of scattering material
along the line of sight and on the wavenumber spectrum of irregularities,
for simplicity we adopt a one-sided exponential form with a characteristic
$e^{-1}$ broadening time that scales as 
$\tau_{\rm iss}(f)=\tau_{\rm 1GHz}(f/{\rm 1\,GHz})^{-4}$.
We modeled the intrinsic profile of each pulsar as a 
sum of between one and five Gaussian components.  The position and width
of each component was held fixed across all four subbands, but the
amplitudes were free to vary.  The fit included scattering time,
$\tau_{\rm 1GHz}$; an arbitrary offset in DM, $\Delta$DM, to account
for biases in the DM used to generated the profiles (since the dedispersion
used for those profiles did not account for scattering); the position and
width of each component; and the amplitude of each component in each subband.
Thus, a total of 8--32 parameters were 
free to vary in the fit for each pulsar, depending on the number of
Gaussian components in the model intrinsic profile.
Some profiles had noisy
baselines, presumably due to radio frequency interference; 
a polynomial (up to 20th order) was included in the model
fit to such profiles.

This model proved satisfactory for all pulsars in our sample.
Examples of fitted pulse profiles are in Figure~\ref{fig:scatter}.
The results of all the fits are given in Table~\ref{table:scatter}, which
lists $\tau_{\rm 1GHz}$; $\Delta$DM; the number of components
in the profile model, $n_{\rm comp}$; and the goodness-of-fit
statistic, $\chi^2_{\nu}$.
Scattering was detectable in a large number of the sources,
and the time scales were determined with precision of a few percent
for the most highly scattered sources.
The fits are characterized by reduced
chi-squared values around $\chi_{\nu}^2 = 1.0\pm 0.5$ for most sources. 
Even though this method fits the pulse profile well, there is an
uncertainty in assuming Gaussian-shaped components in the intrinsic
profile models. For a multi-component  model combined with a small amount
of scattering, the
second or third component is covariant with scattering time. We are not able
to resolve the ambiguity with the current data.  Observations
at lower frequency (and hence higher scattering) and measurements of
scintillation bandwidths could help resolve these covariances.  

We have not tested for departures from the assumed
frequency dependence $\tau_{\rm iss} \propto f^{-4}$ or for
the dependence of the results on alternative forms for the
PBFs, such as those that result from electron-density variations
having a Kolmogorov wavenumber
spectrum and distributed in a thin screen or an extended medium
 \citep{lr99}.
Anisotropic scattering also presents alternative PBF forms as
does any variation of the scattering transverse to the line of
sight \citep[][]{cl01}.

A comparison between
our measured scattering time scales for pulsars in the inner 
Galaxy ($30^\circ<l<80^\circ$) and the predictions
of the NE2001 electron-density model \citep{cl02} is given in
Figure~\ref{fig:scatter_gal}.
The pulse broadening times in our sample of pulsars displays the same
amount of variation at a given value of DM as is seen in other objects
\citep[e.g.,][]{cl02, bcc+04}.   While the NE2001 electron-density model generally
underpredicts the pulse broadening times for the pulsars in our sample with the largest pulse
broadening, the broadening times are consistent with the observed ranges from the larger sample of 
objects used to construct NE2001 and analyzed in \citet[][]{bcc+04}.    The underprediction
is especially notable for objects in the Cygnus direction (as seen at the higher
Galactic longitudes in Figure~\ref{fig:scatter_gal}).   This is to be expected because
the Cygnus superbubble comprises discrete {\sc H\,ii} regions that are known to produce excess scattering \citep[][]{fsm89}.   

All but one of the pulsars (PSR~J2005+3547) have DMs that can be
accommodated by the NE2001 model, but some of the DM values and pulse
broadening times are undoubtedly enhanced by intersection of the line of
sight with particular electron-density enhancements not included in NE2001.
A new electron density model is now under development that will include the
DM values and scattering measurements discussed here.

The scattering measurements imply that searches for millisecond pulsars
along low-Galactic-latitude lines of sight in our observing
region of $32^\circ<\ell<77^\circ$ will be limited by scattering
at very large distances.  
Presuming a scaling law of $\tau_{\rm iss}\propto f^{-4}$, 
the median measured value $\tau_{\rm 1GHz}\approx 4\,{\rm ms}$ 
among our pulsars scales to $\tau_{\rm 1.42GHz}\approx 1\,{\rm ms}$.
Scattering of this magnitude significantly curtails sensitivity
to the fastest-spinning pulsars, especially those with periods of 
order 1~ms.  (It is important to note that there is still
a large volume of space, likely extending to distances of several
kiloparsecs, for which scattering is not significant, and within
which there is promise for a large number of millisecond pulsars
to be detected by surveys such as PALFA.)

\section{X-ray Observations of PSR J1856+0245}\label{sec:1856xray}

The 81-ms pulsar J1856+0245 has the smallest characteristic age,
21~kyr, and largest spin-down luminosity, $\dot{E}=4.6 \times
10^{36}\,{\rm erg\, s}^{-1}$, of the pulsars observed for this paper.
\citet{hng+08} found this pulsar to be coincident with the extended
TeV source HESS J1857+026 and noted faint X-ray emission in the
pulsar's vicinity in archival {\it ASCA} data.  They concluded that
HESS J1857+026 was plausibly a pulsar wind nebula (PWN) associated
with PSR J1856+0245.  \citet{rgv+12} examined 36 months of {\it Fermi}
Large Area Telescope $\gamma$-ray data toward this pulsar.  They
detected emission coincident with the pulsar and extended TeV source,
but they did not detect pulsations.  They interpreted the unpulsed
emission from the region as a putative GeV PWN, and performed
multi-wavelength (keV--TeV) modeling of the spectral energy
distribution.  Here we describe dedicated {\it XMM-Newton} and {\it
  Chandra} X-ray observations of this region, which were obtained to
verify the PWN interpretation.  These observations were used in
\citet{rgv+12} to place a limit on the flux of any extended X-ray
emission.  We present for the first time the X-ray spectrum of
the pulsar and further discuss its association with the GeV--TeV
emission.

\subsection{\textit{XMM-Newton}}
\textit{XMM-Newton} observed the field of PSR J1856+0245 on 2008 March 27--28
(ObsID 0505920101) for a total of 55.3 ks. Data were
acquired from only the European Photon Imaging Camera (EPIC) pn
detector, operating in full-frame imaging mode with the thick optical
filter in place. Unfortunately, due to tests being conducted
simultaneously on the MOS1 and MOS2 detectors, no data are available
from those instruments. The EPIC pn field of view was centered
close to the timing position of PSR J1856+0245 and also included the
centroid of the TeV nebula HESS J1857+026 (Figure~\ref{fig:1856chandra}).
The raw (ODF) pn data were
reprocessed using the {\sc Epchain} pipeline in {\sc SAS}.\footnote{The
  \textit{XMM-Newton} SAS is developed and maintained by the Science
  Operations Centre at the European Space Astronomy Centre and the
  Survey Science Centre at the University of Leicester.} 
The data were
subsequently screened for instances of strong soft-proton flaring.
The filtering procedure left 30.2 ks of useful exposure time. 
The Reflection Grating Spectrometer data provided no useful
source information and are not included in our analysis.
The {\sc SAS} source detection task {\sc edetect\_chain} found a source
coincident with the radio timing position of PSR J1856+0245, XMMU
J185650.8+024545.

Photons were extracted from a circular region of radius $10^{\prime\prime}$ centered
on the pulsar, which encloses $\sim$52\% of the 
total energy.  This
relatively small extraction radius was chosen to minimize
contamination from the nearby ($\sim 22^{\prime\prime}$) source XMMU
J185651.8+024530.  An estimate of the background was obtained from
several source-free regions around the pulsar position. The spectral
analysis was performed using {\sc Xspec}\footnote{\url{http://heasarc.nasa.gov/docs/xanadu/xspec/index.html}.}
12.5.0.5.  The source spectrum was grouped to ensure at least 15 photons
per energy bin. 

The X-ray counterpart of PSR J1856+0245 shows a hard
spectrum, most likely because it is distant and heavily absorbed. We
fit the spectrum of the source with both absorbed power-law and
thermal models. Given the low photon count, both models describe the
data adequately well, though an absorbed blackbody model gives a
temperature of $kT\approx1.5-3\,{\rm keV}$, too hot to be the
surface temperature of the neutron star. For this reason, the power law
model is preferred.  For young pulsars, such emission is typically due
to particle acceleration in the pulsar magnetosphere. The best-fit
parameters for this model are only weakly constrained due to the
limited number of photons with absorption
$N_{\rm H}=(4.9^{+3.2}_{-2.4})\times 10^{22}\,{\rm cm}^{-2}$ and
photon index $\Gamma=1.2^{+0.8}_{-0.4}$. The implied unabsorbed flux is
$\sim$$8.3\times10^{-14}$ erg cm$^{-2}$ s$^{-1}$ (2--10 keV), which
for a distance of $9$~kpc (Table~\ref{table:scatter}) corresponds to a luminosity of
$L_X\sim8\times10^{32}$ erg s$^{-1}$.

The 73.4 ms time resolution of the EPIC pn full-frame exposure and 
the 3.2 s frame time of the ACIS-I observation preclude the search 
for X-ray pulsations from PSR J1856+0245.

\subsection{\textit{Chandra}}

PSR J1856+0245 was observed with \textit{Chandra} ACIS-I for
39~ks on 2011 February 28 (ObsID 12557). These data were recorded in
VFAINT and Timed Exposure modes and were analyzed using
{\sc CIAO}\footnote{Chandra Interactive Analysis of Observations 
\citep{fma+06},
\url{http://cxc.harvard.edu/ciao/}} version 4.3.1 with {\sc
  CALDB}~4.4.3.  PSR J1856+0245 is clearly detected as a point source
(Figure~\ref{fig:1856chandra}).  As also discussed in \citet{rgv+12},
there is no evidence for extended emission surrounding the
pulsar. Subtracting a simulated point spread function using {\sc
  ChaRT}\footnote{The Chandra Ray Tracer,
  \url{http://cxc.harvard.edu/soft/ChaRT/cgi-bin/www-saosac.cgi}} and
{\sc marx}\footnote{\url{http://space.mit.edu/cxc/marx/index.html}.}
from the observed point source yields no excess counts that could
arise due to a compact PWN. Extracting the emission from an annular
region with inner and outer radii of 2$''$ and 15$''$, respectively,
we find an upper limit on the unabsorbed flux of $5\times10^{14}$ erg
s$^{-1}$ cm$^{-2}$ (1--10 keV, 3$\sigma$ confidence), corresponding
to a luminosity of $\lesssim$$5 \times 10^{32}$ erg s$^{-1}$. In
deriving this limit we assumed a power-law index of $\Gamma =
1.5$ and an equivalent neutral hydrogen column depth $N_{\rm H} = 4
\times 10^{22}\,{\rm cm}^{-2}$.  The latter is somewhat higher than
would be predicted by a simple empirical correlation between $N_{\rm
  H}$ and DM \citep{hnk13}, so can be regarded as conservative.  Note
that the outer extraction radius was chosen based on the X-ray PWN for
pulsars with comparable $\dot{E}$ by scaling their angular size with
the distance \citep{kp08}, assuming a distance of 9 kpc
(Table~\ref{table:scatter}).

\subsection{Association between PSR J1856+0245 and HESS J1857+026}

\citet{hng+08} showed that PSR J1856+0245 is coincident with the faint
      {\it ASCA} X-ray source AX J185651+0245 and suggested that the
      latter might represent an X-ray PWN associated with the pulsar.
      Our {\it XMM-Newton} and {\it Chandra} observations confirm
      X-ray emission from the pulsar itself, but find no evidence of
      surrounding diffuse emission.  While the discovery of an X-ray
      PWN---especially one which shows a morphology indicative of an
      association with the TeV PWN \citep[e.g.,][]{aab+05}---would
      strengthen the link between PSR J1856+0245 and HESS J1857+026,
      the lack of an observable X-ray PWN does not eliminate the possibility of
      association.  Given the distance of 9\,kpc inferred from the pulsar's
      DM, it is possible that
      the X-ray PWN is simply too faint to be detected in the
      current data.  Indeed, \citet{rgv+12} show that it is possible
      to derive a self-consistent model of the spectral energy
      distribution that accounts for the low X-ray nebula flux.

We have taken a critical look at other X-ray sources detected in our
{\it XMM-Newton} and {\it Chandra} observations, and find no evidence
for extended sources of a possible PWN nature (Figure~\ref{fig:1856chandra}).  These
observations do not cover the full extent of the TeV nebula, but they
do include the source centroid.  Furthermore, the PALFA survey has
covered the region of the HESS TeV emission and as such any
additional, undiscovered, young pulsar in the region would have to be
quite faint, $S_{1400} \lesssim 0.2$\,mJy (depending on the duty cycle,
DM, and scattering).  We thus conclude that PSR
J1856+0245 remains the most viable association for HESS J1857+026.

\section{Discussion}\label{sec:discussion}

We have presented timing properties and pulsar profile morphologies 
of 35
pulsars discovered in the PALFA survey.  Due to this survey's depth and
its sensitivity to high-DM pulsars, these pulsars are relatively distant,
with a median distance of 7.1\,kpc 
compared to the median distance of 4.2\,kpc
of Galactic pulsars listed in the ATNF pulsar catalog
(Figure~\ref{fig:galplot} and Table~\ref{table:scatter}).

Despite their large distances, the properties of these pulsars
are similar to those of the previously known non-recycled Galactic pulsar
population.  We used Kolmogorov-Smirnov tests to compare the distributions
of spin-down ages, magnetic field strengths, and spin-down energies 
of the pulsars in this paper with those of the known pulsar population.
We excluded recycled pulsars from this comparison by imposing a cutoff
$\dot{P}>10^{-17}$.
We found no significant differences between the two populations.
(A more thorough analysis of the population
of pulsars discovered by PALFA, using more powerful statistical techniques 
and considering the details of the telescope search pointings, is beyond
the scope of the present paper.)

Four of the sources, PSRs J1856+0245, J1909+0749, J1928+1746 and J1934+2352,
have spin-down ages $t_s<10^5$\,yr.  Pulsars of such young age are
often found in supernova remnants.   However, Green's Supernova Remnant
catalog \citep{gre09} lists no remnants coincident with the positions of
these pulsars.  The remnant catalog is known to be incomplete, however, and
dedicated radio interferometric observations of these objects in particular
to look for remnants could be fruitful.

Young and energetic radio pulsars often have counterparts at other wavelengths,
such as the X-ray source at the position of PSR~J1856+0245 described above.  We have
checked the {\it ROSAT} all-sky survey for X-ray point sources at the
positions of all the pulsars having $\dot{E} > 2 \times
10^{35}$~erg~s$^{-1}$, but we have found no counterparts.

\begin{figure*}[p]
\begin{center}
\includegraphics[width=6.0in]{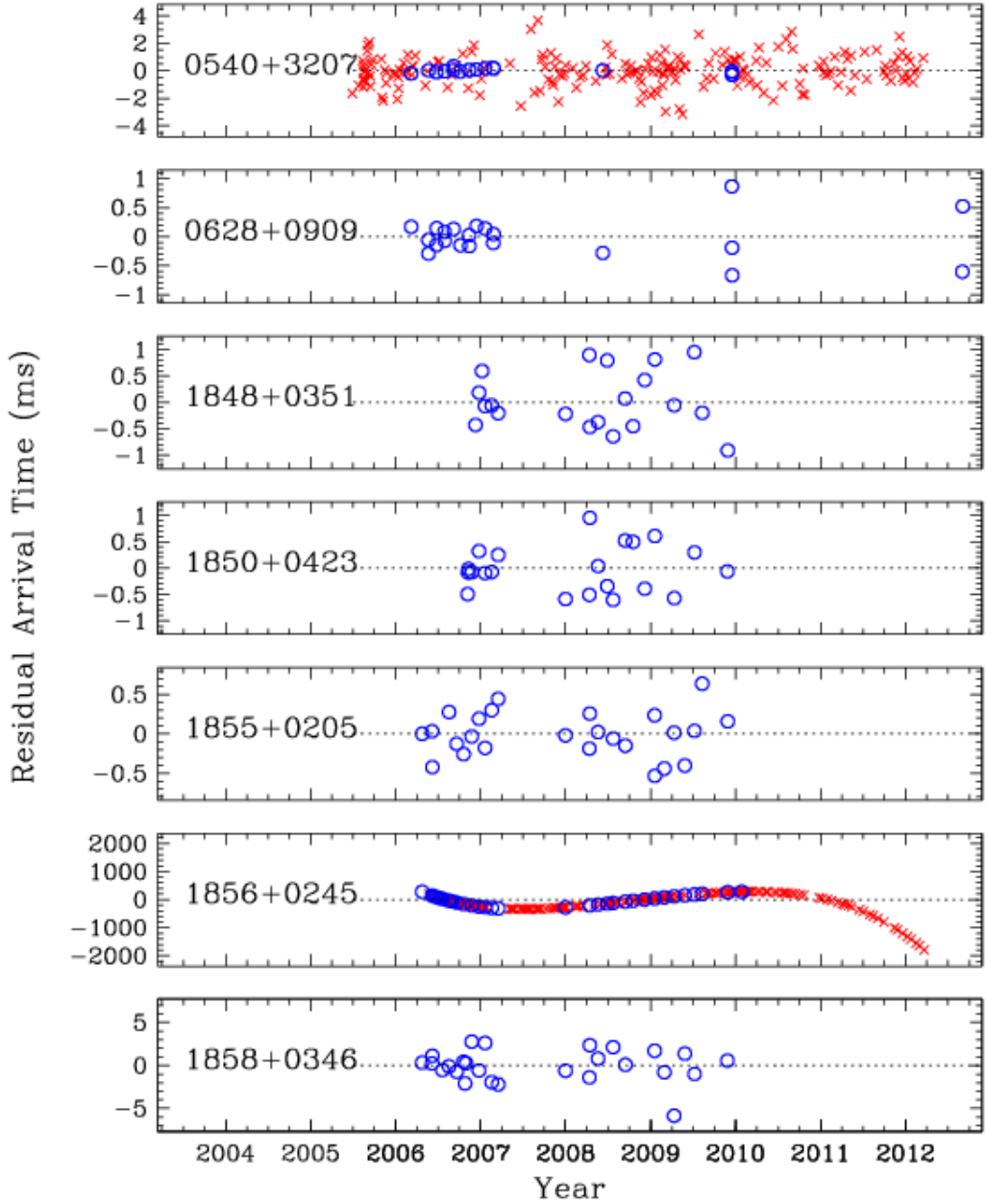}
\caption{Part (a).  \label{fig:resid1}Residual pulse arrival times of all 
pulsars in this paper
are given in (a)-(e).  The 
points are daily average residuals at Arecibo 
(blue circles)
and Jodrell Bank 
(red $\times$'s).
For each pulsar, the residual
plot was made by performing
a timing fit, possibly including whitening and glitch parameters
as described in the text, to find the best-fit parameters; 
freezing the position at its best-fit value; removing whitening
and glitches from the model; re-fitting for spin period and spin-down
rate; and averaging the resulting residual pulse arrival times
within each day at each observatory.  
}
\end{center}
\end{figure*}

As discussed in Section~\ref{sec:timing},
many of the pulsars in this paper show deviations from steady spin-down
(Figures~\ref{fig:resid1}(a)--(e)).
The spin evolution of a pulsar is conventionally
quantified by a braking index, $n$, defined through $\dot{\nu}\propto\nu^n$,
where $\nu=1/P$ is the spin frequency and $\dot{\nu}$ is its time derivative.
For a pulsar undergoing energy loss due to magnetic dipole
rotation, $n=3$.  In principle, the braking index for a pulsar can be calculated
from $n=\nu\ddot{\nu}/\dot{\nu}^2$ if the
second time derivative of spin frequency, $\ddot{\nu}$, is known
\citep{lk05}.  In practice, such measurements have been achieved for only
a few pulsars
\citep[e.g.,][and references therein]{lkgm05}.  Most measurements of $\ddot{\nu}$
and inferred values of $n$ have been much larger in magnitude than predicted by steady 
spin-down models.  
\cite{hlk10}\ analyzed
long-term pulsar timing data of hundreds of pulsars and found that 
pulsars with spin-down ages less than $10^{5}$\,yr all had positive values
of $n$, whereas older pulsars had a mix of positive and
negative values.  They interpreted this as evidence that young pulsars are
recovering from glitches that occurred prior to the starts of the data sets,
while older pulsars are dominated by stochastic processes.

Significant values of $\ddot{\nu}$, and hence $n$, can be measured for
about half of our sources (all those for which $n_{\rm fit}>1$ in Table~\ref{table:parameters}).
The four pulsars with spin-down
ages less than $10^5$\,yr  all show positive values of $n$:
$n=24.5, 11.3, 22.9, {\rm and}\ 12.3$ for PSRs J1856+0245, J1909+0749, 
J1928+1746, and J1934+2352, respectively, consistent with the findings of
\cite{hlk10}.    The older pulsars show a mix of positive and negative values of $n$.
All measured values of $n$ are sufficiently large in magnitude to suggest
that they are not associated with steady spin-down of these pulsars.

The dispersion measurements and scattering measurements presented in 
Sections~\ref{sec:timing} and \ref{sec:scattering} are
a promising start of the detailed study of the ionized interstellar medium
in this range of Galactic longitudes.  The scattering measurements could
be greatly enhanced by collecting and analyzing profiles of these sources at longer
wavelengths.  

The 35 pulsars in this paper are just the tip of the iceberg.
As of this writing, the ongoing PALFA survey had discovered
more than 100 pulsars.  Study of these objects will enhance our
knowledge of the pulsar population and the interstellar medium
in this sector of the Galaxy.

\acknowledgements 

The Arecibo Observatory is operated by SRI International under a
cooperative agreement with the National Science Foundation (AST-1100968),
and in alliance with Ana G. M\'endez-Universidad Metropolitana, and the
Universities Space Research Association.  
The scientific results reported
in this article are based in part on observations made by the {\it Chandra X-ray
Observatory}  and of software provided by the Chandra X-Ray Center (CXC) in
the application package CIAO.  This work is based on observations obtained
with {\it XMM-Newton}, an ESA science mission with instruments and contributions
directly funded by ESA Member States and NASA. 

This work was supported by National Science Foundation grants 0555655,
0647820, 0807151, 1104902, 1105572, and 1148523.  This work was supported
by the Max Plank Gesellschaft.

P.~F.\ acknowledges financial support by the European Research Council for the ERC
Starting Grant BEACON under contract 279702.  
J.~W.~T.~H. and J.~v.~L.\ acknowledge support
from the Netherlands Foundation for Scientific Research (NWO).
V.~M.~K.\ was supported
by an NSERC Discovery Grant, the Canadian Institute for Advanced Research,
a Canada Research Chair, Fonds de Recherche Nature et Technologies, and the
Lorne Trottier Chair in Astrophysics.  
B.~K.\ acknowledges the support of the Max Planck Society.  
P.~L.\ acknowledges the support of IMPRS Bonn/Cologne and NSERC PGS-D.
D.~R.~L.\ and M.~A.~M.\ acknowledge support from West Virginia EPSCoR and the
Research Corporation for Scientific Advancement.
I.~H.~S.\ acknowledges support from an NSERC Discovery
Grant and Discovery Accelerator Supplement.

We thank R. Rousseau for supplying the {\it Fermi}-LAT contours
in Figure~\ref{fig:1856chandra}.


\begin{deluxetable}{ccllclllc}
\tabletypesize{\scriptsize}
\tablecaption{\label{table:parameters}Timing Parameters of 35 Pulsars\tablenotemark{a}}
\tablehead{%
PSR & Previous        & \multicolumn{1}{c}{$\alpha$       } & \multicolumn{1}{c}{$\delta$       } & \multicolumn{1}{c}{Epoch of $P$      } & \multicolumn{1}{c}{$P$    }  &   \multicolumn{1}{c}{$\dot{P}$}  &  \multicolumn{1}{c}{DM                } & $n_{\rm fit}$ \\
    & Reference\tablenotemark{b} & \multicolumn{1}{c}{(J2000)         } & \multicolumn{1}{c}{(J2000)         } & \multicolumn{1}{c}{(MJD)             } & \multicolumn{1}{c}{(s)    }  &   \multicolumn{1}{c}{($10^{-15}$)}  &  \multicolumn{1}{c}{   (pc\,cm$^{-3}$)} &                
}
\startdata
J0540+3207  &    1 &  05:40:37.116(1)  &  32:07:37.3(1)  &  54780.000  &  0.5242708632083(5)  &  \phn \phn 0.44811(2)  &  \phn \phn 61.97(4)  &  1  \\
J0628+0909  &  1,2 &  06:28:36.183(5)  &  09:09:13.9(3)  &  54990.000  &  1.241421391299(3)  &  \phn \phn 0.5479(2)  &  \phn \phn 88.3(2)  &  1  \\
J1848+0351  &      &  18:48:42.24(1)  &  03:51:35.7(3)  &  54620.000  &  0.191442663494(2)  &  \phn \phn 0.0673(1)  &  \phn \phn 336.6(4)  &  1  \\
J1850+0423  &      &  18:50:23.43(1)  &  04:23:09.2(4)  &  54600.000  &  0.290716217162(3)  &  \phn \phn 0.0914(2)  &  \phn \phn 265.8(4)  &  1  \\
J1855+0205  &      &  18:55:42.046(5)  &  02:05:36.4(2)  &  54510.000  &  0.2468167699732(9)  &  \phn \phn 0.06466(7)  &  \phn \phn 867.3(2)  &  1  \\ [2pt]
J1856+0245  &    3 &  18:56:50.9(3)  &  02:45:47(9)  &  54930.000  &  0.08090668906(4)  &  \phn \phn 62.117(4)  &  \phn \phn 623.5(2)  &  N  \\
J1858+0346  &      &  18:58:22.36(2)  &  03:46:37.8(8)  &  54510.000  &  0.256843797950(5)  &  \phn \phn 2.0401(4)  &  \phn \phn 386(1)  &  1  \\
J1904+0738  &    1 &  19:04:07.533(1)  &  07:38:51.69(4)  &  54760.000  &  0.2089583321216(2)  &  \phn \phn 0.410898(7)  &  \phn \phn 278.32(8)  &  1  \\
J1905+0902  &    1 &  19:05:19.535(2)  &  09:02:32.49(8)  &  54570.000  &  0.2182529126846(9)  &  \phn \phn 3.49853(8)  &  \phn \phn 433.4(1)  &  3  \\
J1909+0641  &    2 &  19:09:29.052(4)  &  06:41:25.8(2)  &  54870.000  &  0.741761952452(6)  &  \phn \phn 3.2239(7)  &  \phn \phn 36.7(2)  &  1  \\ [2pt]
J1909+0749  &      &  19:09:08.2(2)  &  07:49:32(5)  &  54870.000  &  0.23716129322(9)  &  \phn \phn 151.920(9)  &  \phn \phn 539.36(5)  &  N  \\
J1916+1225  &      &  19:16:20.045(1)  &  12:25:53.94(4)  &  54570.000  &  0.227387488792(2)  &  \phn \phn 23.4516(2)  &  \phn \phn 265.31(3)  &  8  \\
J1917+1737  &      &  19:17:23.976(5)  &  17:37:31.6(1)  &  55070.000  &  0.334725226760(4)  &  \phn \phn 0.324(1)  &  \phn \phn 208.0(2)  &  1  \\
J1919+1314  &      &  19:19:32.986(5)  &  13:14:37.3(1)  &  54790.000  &  0.571399860595(2)  &  \phn \phn 3.78771(6)  &  \phn \phn 613.4(2)  &  1  \\
J1919+1745  &    2 &  19:19:43.342(4)  &  17:45:03.79(8)  &  55320.000  &  2.081343459724(9)  &  \phn \phn 1.7050(4)  &  \phn \phn 142.3(2)  &  1  \\ [2pt]
J1924+1631  &      &  19:24:54.88(3)  &  16:31:48.8(6)  &  54690.000  &  2.9351864592(2)  &  \phn \phn 364.212(4)  &  \phn \phn 518.5(9)  &  2  \\
J1928+1746  &    1 &  19:28:42.553(1)  &  17:46:29.62(3)  &  54770.000  &  0.0687304058253(2)  &  \phn \phn 13.190700(6)  &  \phn \phn 176.68(5)  &  4  \\
J1934+2352  &      &  19:34:46.19(2)  &  23:52:55.9(3)  &  54650.000  &  0.17843152366(2)  &  \phn \phn 130.699(2)  &  \phn \phn 355.5(2)  &  8  \\
J1938+2010  &      &  19:38:08.34(2)  &  20:10:51.7(4)  &  54940.000  &  0.68708185664(2)  &  \phn \phn 3.399(1)  &  \phn \phn 327.7(8)  &  4  \\
J1940+2337  &      &  19:40:35.486(7)  &  23:37:46.5(2)  &  54560.000  &  0.546824157964(7)  &  \phn \phn 76.7965(2)  &  \phn \phn 252.1(3)  &  2  \\ [2pt]
J1941+2525  &      &  19:41:20.80(1)  &  25:25:05.3(2)  &  54920.000  &  2.30615269700(2)  &  \phn \phn 160.8349(5)  &  \phn \phn 314.4(4)  &  1  \\
J1946+2535  &      &  19:46:49.10(5)  &  25:35:51.5(9)  &  54810.000  &  0.51516705131(6)  &  \phn \phn 5.641(5)  &  \phn \phn 248.81(4)  &  N  \\
J1947+1957  &      &  19:47:19.435(4)  &  19:57:08.39(9)  &  55190.000  &  0.157508542541(4)  &  \phn \phn 0.5226(1)  &  \phn \phn 185.8(2)  &  1  \\
J1948+2551  &      &  19:48:17.579(1)  &  25:51:51.95(2)  &  54930.000  &  0.1966268136138(4)  &  \phn \phn 9.02244(2)  &  \phn \phn 289.27(5)  &  3  \\
J1949+2306  &      &  19:49:07.32(1)  &  23:06:55.5(2)  &  54910.000  &  1.31937338530(1)  &  \phn \phn 0.1231(3)  &  \phn \phn 196.3(5)  &  1  \\ [2pt]
J1953+2732  &      &  19:53:07.80(2)  &  27:32:48.2(4)  &  54570.000  &  1.33396499366(2)  &  \phn \phn 1.773(1)  &  \phn \phn 195.4(9)  &  1  \\
J2005+3547  &      &  20:05:17.493(9)  &  35:47:25.4(1)  &  54540.000  &  0.615033894871(4)  &  \phn \phn 0.2808(3)  &  \phn \phn 401.6(3)  &  1  \\
J2006+3102  &      &  20:06:11.0(3)  &  31:02:03(4)  &  54800.000  &  0.16369523645(9)  &  \phn \phn 24.872(9)  &  \phn \phn 107.16(1)  &  N  \\
J2007+3120  &      &  20:07:09.012(6)  &  31:20:51.54(9)  &  54920.000  &  0.60820531457(1)  &  \phn \phn 15.604(1)  &  \phn \phn 191.5(2)  &  9  \\
J2009+3326  &    1 &  20:09:49.16(2)  &  33:26:10.2(3)  &  54430.000  &  1.43836860189(2)  &  \phn \phn 1.468(2)  &  \phn \phn 263.6(8)  &  1  \\ [2pt]
J2010+2845  &      &  20:10:05.068(2)  &  28:45:29.17(3)  &  54810.000  &  0.5653693451984(6)  &  \phn \phn 0.09072(2)  &  \phn \phn 112.47(8)  &  1  \\
J2010+3230  &    1 &  20:10:26.50(1)  &  32:30:07.3(2)  &  54430.000  &  1.44244752060(1)  &  \phn \phn 3.616(1)  &  \phn \phn 371.8(5)  &  1  \\
J2011+3331  &    1 &  20:11:04.926(2)  &  33:31:24.64(3)  &  54490.000  &  0.931733093470(4)  &  \phn \phn 1.7857(1)  &  \phn \phn 298.58(6)  &  2  \\
J2013+3058  &      &  20:13:34.253(3)  &  30:58:50.69(5)  &  54600.000  &  0.2760276934460(5)  &  \phn \phn 0.15194(4)  &  \phn \phn 148.7(1)  &  1  \\
J2018+3431  &    1 &  20:18:53.196(2)  &  34:31:00.51(2)  &  54760.000  &  0.3876640866385(8)  &  \phn \phn 1.83649(2)  &  \phn \phn 222.35(7)  &  2  \\
\enddata
\tablenotetext{a}{Figures in parentheses are uncertainties in the last digit quoted.}
\tablenotetext{b}{References:  1. \citet{cfl+06}   2. \citet{dcm+09}   3. \citet{hng+08} .} 
\end{deluxetable}

\begin{deluxetable}{cccc}
\tabletypesize{\scriptsize}
\tablecaption{\label{table:derived}Quantities Derived from Rotation Parameters of Table~\ref{table:parameters}.}
\tablehead{PSR & $\log\,t_{\rm s}$  &  $\log\,B$  &  $\log\,\dot{E}$  \\
    & (log\,yr)   & (log\,G)   & (log\,erg\,s$^{-1}$) }
\startdata
J0540+3207 & 7.27 & 11.7 & 32.1 \\
J0628+0909 & 7.56 & 11.9 & 31.0 \\
J1848+0351 & 7.65 & 11.1 & 32.6 \\
J1850+0423 & 7.70 & 11.2 & 32.2 \\
J1855+0205 & 7.78 & 11.1 & 32.2 \\ [2pt]
J1856+0245 & 4.31 & 12.4 & 36.7 \\
J1858+0346 & 6.30 & 11.9 & 33.7 \\
J1904+0738 & 6.91 & 11.5 & 33.2 \\
J1905+0902 & 5.99 & 11.9 & 34.1 \\
J1909+0641 & 6.56 & 12.2 & 32.5 \\ [2pt]
J1909+0749 & 4.39 & 12.8 & 35.6 \\
J1916+1225 & 5.19 & 12.4 & 34.9 \\
J1917+1737 & 7.21 & 11.5 & 32.5 \\
J1919+1314 & 6.38 & 12.2 & 32.9 \\
J1919+1745 & 7.29 & 12.3 & 30.9 \\ [2pt]
J1924+1631 & 5.11 & 13.5 & 32.7 \\
J1928+1746 & 4.92 & 12.0 & 36.2 \\
J1934+2352 & 4.34 & 12.7 & 36.0 \\
J1938+2010 & 6.51 & 12.2 & 32.6 \\
J1940+2337 & 5.05 & 12.8 & 34.3 \\ [2pt]
J1941+2525 & 5.36 & 13.3 & 32.7 \\
J1946+2535 & 6.16 & 12.2 & 33.2 \\
J1947+1957 & 6.68 & 11.5 & 33.7 \\
J1948+2551 & 5.54 & 12.1 & 34.7 \\
J1949+2306 & 8.23 & 11.6 & 30.3 \\ [2pt]
J1953+2732 & 7.08 & 12.2 & 31.5 \\
J2005+3547 & 7.54 & 11.6 & 31.7 \\
J2006+3102 & 5.02 & 12.3 & 35.3 \\
J2007+3120 & 5.79 & 12.5 & 33.4 \\
J2009+3326 & 7.19 & 12.2 & 31.3 \\ [2pt]
J2010+2845 & 7.99 & 11.4 & 31.3 \\
J2010+3230 & 6.80 & 12.4 & 31.7 \\
J2011+3331 & 6.92 & 12.1 & 31.9 \\
J2013+3058 & 7.46 & 11.3 & 32.4 \\
J2018+3431 & 6.52 & 11.9 & 33.1 \\
\enddata
\end{deluxetable}

\begin{deluxetable}{crrccccccccc}
\tabletypesize{\scriptsize}
\tablecaption{\label{table:scatter}Galactic Positions, Distance, and Scattering Properties.
}
\tablehead{
 & & & && \multicolumn{2}{c}{NE2001 Predictions\tablenotemark{a}} && \multicolumn{4}{c}{Measured Values\tablenotemark{b}} \\ 
\cline{6-7}\cline{9-12}
PSR & \multicolumn{1}{c}{$l$}         & \multicolumn{1}{c}{$b$}         & \multicolumn{1}{c}{DM}               && \multicolumn{1}{c}{$d$}   & \multicolumn{1}{c}{$\tau_{\rm 1GHz}$} &&  $\tau_{\rm 1GHz}$      & $\Delta$DM      & $n_{\rm comp}$ & $\chi^2_{\nu}$   \\
    & \multicolumn{1}{c}{$(^\circ)$} & \multicolumn{1}{c}{$(^\circ)$} & \multicolumn{1}{c}{(pc\,cm$^{-3}$)} && \multicolumn{1}{c}{(kpc)} & \multicolumn{1}{c}{(ms)}                && \multicolumn{1}{c}{(ms)} & (pc\,cm$^{-3}$) &       
}
\startdata
J1924+1631 &   51.405 & $    0.318 $ &    518.5 && \phantom{$>$}14.0  & {\phn}0.50 & & 49(2){\phd\phn\phn} & {\phn}$-10.4(8)${\phn}{\phn} & 2 &  1.3 \\
J1858+0346 &   37.084 & $    0.182 $ &    386.6 && \phantom{$>$0}7.0  & {\phn}4.97 & & 42(3){\phd\phn\phn} & {\phn}$-3.0(4)${\phn}{\phn} & 3 &  1.1 \\
J2010+3230 &   70.391 & $   -0.495 $ &    371.8 && \phantom{$>$}12.0  & {\phn}0.08 & & 24.8(6){\phn} & {\phn}$-3.8(1)${\phn}{\phn} & 1 &  1.0 \\
J1953+2732 &   64.205 & $    0.059 $ &    195.4 && \phantom{$>$0}7.0  & {\phn}0.01 & & 21(3){\phd\phn\phn} & {\phn}$-2.8(2)${\phn}{\phn} & 3 &  1.1 \\
J1919+1314 &   47.895 & $   -0.088 $ &    613.5 && \phantom{$>$}15.2  & {\phn}1.87 & & 18.7(9){\phn} & {\phn}$-2.11(5)${\phn} & 2 &  1.5 \\ [2pt]
J1856+0245 &   36.008 & $    0.057 $ &    623.5 && \phantom{$>$0}9.0  & 23.31 & & 18.0(6){\phn} & {\phn}$-3.0(1)${\phn}{\phn} & 2 &  1.0 \\
J1855+0205 &   35.281 & $    0.007 $ &    867.3 && \phantom{$>$}11.6  & 57.33 & & 16.3(9){\phn} & {\phn}$-2.1(1)${\phn}{\phn} & 2 &  1.4 \\
J2005+3547 &   72.581 & $    2.180 $ &    401.7 && $>$50.0  & {\phn}0.07 & & 11(1){\phd\phn\phn} & {\phn}$-1.85(9)${\phn} & 2 &  1.5 \\
J1850+0423 &   36.718 & $    2.231 $ &    265.9 && \phantom{$>$0}6.4  & {\phn}0.28 & & {\phn}7.8(8){\phn} & {\phs}{\phn}$0.0(2)${\phn}{\phn} & 2 &  1.3 \\
J1941+2525 &   61.037 & $    1.263 $ &    314.4 && \phantom{$>$0}9.8  & {\phn}0.03 & & {\phn}7.4(6){\phn} & {\phn}$-2.4(2)${\phn}{\phn} & 2 &  1.5 \\ [2pt]
J1948+2551 &   62.207 & $    0.131 $ &    289.3 && \phantom{$>$0}8.8  & {\phn}0.02 & & {\phn}6.3(2){\phn} & {\phn}$-0.83(1)${\phn} & 4 &  1.5 \\
J1949+2306 &   59.931 & $   -1.423 $ &    196.4 && \phantom{$>$0}7.0  & {\phn}0.01 & & {\phn}5.4(7){\phn} & {\phn}$-1.0(1)${\phn}{\phn} & 3 &  1.4 \\
J1916+1225 &   46.811 & $    0.225 $ &    265.3 && \phantom{$>$0}7.1  & {\phn}0.59 & & {\phn}5.3(2){\phn} & {\phn}$-0.155(9)$ & 4 &  1.6 \\
J1946+2535 &   61.809 & $    0.283 $ &    248.8 && \phantom{$>$0}8.1  & {\phn}0.01 & & {\phn}4.54(7) & {\phn}$-0.97(2)${\phn} & 3 &  1.8 \\
J1938+2010 &   56.115 & $   -0.672 $ &    327.7 && \phantom{$>$0}9.5  & {\phn}0.05 & & {\phn}4(1){\phd\phn\phn} & {\phs}{\phn}$0.1(4)${\phn}{\phn} & 2 &  1.1 \\ [2pt]
J2011+3331 &   71.320 & $   -0.049 $ &    298.6 && \phantom{$>$0}8.8  & {\phn}4.16 & & {\phn}4.39(9) & {\phn}$-0.84(2)${\phn} & 3 &  1.9 \\
J1940+2337 &   59.397 & $    0.529 $ &    252.2 && \phantom{$>$0}8.2  & {\phn}0.02 & & {\phn}4.2(5){\phn} & {\phn}$-0.8(1)${\phn}{\phn} & 2 &  1.3 \\
J2007+3120 &   69.042 & $   -0.535 $ &    191.6 && \phantom{$>$0}6.8  & {\phn}0.01 & & {\phn}3.8(8){\phn} & {\phn}$-0.52(6)${\phn} & 2 &  1.6 \\
J1919+1745 &   51.897 & $    1.987 $ &    142.3 && \phantom{$>$0}5.2  & {\phn}0.01 & & {\phn}2.9(2){\phn} & {\phn}$-0.60(5)${\phn} & 4 &  1.8 \\
J1909+0749 &   41.909 & $   -0.340 $ &    539.4 && \phantom{$>$0}9.5  & {\phn}6.31 & & {\phn}1.9(1){\phn} & {\phn}$-0.32(4)${\phn} & 5 &  1.5 \\ [2pt]
J1904+0738 &   41.180 & $    0.680 $ &    278.3 && \phantom{$>$0}6.3  & {\phn}1.01 & & {\phn}1.61(7) & {\phn}$-0.30(2)${\phn} & 3 &  1.4 \\
J2018+3431 &   73.044 & $   -0.841 $ &    222.3 && \phantom{$>$0}7.2  & {\phn}0.02 & & {\phn}1.2(1){\phn} & {\phn}$-0.35(3)${\phn} & 2 &  1.3 \\
J1905+0902 &   42.555 & $    1.056 $ &    433.5 && \phantom{$>$0}9.1  & {\phn}1.78 & & {\phn}0.9(2){\phn} & {\phn}$-0.22(6)${\phn} & 3 &  1.6 \\
J2013+3058 &   69.485 & $   -1.886 $ &    148.8 && \phantom{$>$0}6.0  & {\phn}0.00 & & $<$0.9{\phantom{000)}} & {\phn}$-0.10(6)${\phn} & 3 &  1.6 \\
J1917+1737 &   51.528 & $    2.418 $ &    208.1 && \phantom{$>$0}7.0  & {\phn}0.02 & & $<$1.0{\phantom{000)}} & {\phn}$-0.1(3)${\phn}{\phn} & 2 &  1.2 \\ [2pt]
J1928+1746 &   52.931 & $    0.114 $ &    176.7 && \phantom{$>$0}5.8  & {\phn}0.04 & & $<$0.6{\phantom{000)}} & {\phn}$-0.11(4)${\phn} & 5 &  1.3 \\
J1848+0351 &   36.058 & $    2.366 $ &    336.7 && \phantom{$>$0}7.9  & {\phn}0.41 & & $<$1.0{\phantom{000)}} & {\phs}{\phn}$0(1)${\phd\phn\phn\phn} & 1 &  1.5 \\
J2006+3102 &   68.667 & $   -0.530 $ &    107.2 && \phantom{$>$0}4.7  & {\phn}0.00 & & $<$1.0{\phantom{000)}} & {\phs}{\phn}$0.03(6)${\phn} & 5 &  1.4 \\
J2009+3326 &   71.103 & $    0.124 $ &    263.6 && \phantom{$>$0}7.9  & {\phn}1.73 & & $<$1.0{\phantom{000)}} & {\phs}{\phn}$0.1(1)${\phn}{\phn} & 3 &  1.3 \\
J1934+2352 &   58.966 & $    1.814 $ &    355.5 && \phantom{$>$}11.6  & {\phn}0.03 & & $<$1.0{\phantom{000)}} & {\phs}{\phn}$0.1(2)${\phn}{\phn} & 2 &  1.6 \\ [2pt]
J2010+2845 &   67.210 & $   -2.472 $ &    112.5 && \phantom{$>$0}5.0  & {\phn}0.00 & & $<$1.0{\phantom{000)}} & {\phn}$-0.07(3)${\phn} & 3 &  3.5 \\
J1909+0641 &   40.941 & $   -0.940 $ &     36.7 && \phantom{$>$0}2.2  & {\phn}0.00 & & $<$1.0{\phantom{000)}} & {\phs}{\phn}$0.00(3)${\phn} & 3 &  1.6 \\
J0540+3207 &  176.719 & $    0.761 $ &     62.0 && \phantom{$>$0}1.8  & {\phn}0.00 & & $<$1.0{\phantom{000)}} & {\phn}$-0.060(5)$ & 2 &  2.6 \\
J0628+0909 &  202.190 & $   -0.851 $ &     88.3 && \phantom{$>$0}2.5  & {\phn}0.00 & & $<$1.0{\phantom{000)}} & {\phs}{\phn}$0.144(9)$ & 1 &  3.7 \\
J1947+1957 &   56.987 & $   -2.658 $ &    185.9 && \phantom{$>$0}6.8  & {\phn}0.01 & & $<$1.0{\phantom{000)}} & {\phn}$-0.03(4)${\phn} & 2 &  1.5 \\
\enddata
\tablenotetext{a}{Values predicted based on $l$, $b$, and DM, using the NE2001 electron density model of \cite{cl02}.}
\tablenotetext{b}{Values measured from fits to pulsar profiles (Section~\ref{sec:scattering}).  Figures in parentheses are uncertainties in the last digit quoted.}
\end{deluxetable}

\begin{deluxetable}{cllrrr}
\tabletypesize{\scriptsize}
\tablecaption{\label{table:flux}
Flux Densities, Spectral Indices, and Luminosities.}
\tablehead{PSR & \multicolumn{1}{c}{$S_{1400}$} & \multicolumn{1}{c}{$\alpha$} & \multicolumn{1}{c}{$\chi^2_\nu$} &  \multicolumn{1}{c}{$L_{1400}$   } \\
    & \multicolumn{1}{c}{(mJy)     } &                                &                                    &  \multicolumn{1}{c}{(Jy\,kpc$^2$)} 
}
\startdata
J0540+3207 & 0.34(2) &  $-2.0(3)$  & 0.384 &    1.1\ \ \ \   \\
J0628+0909 & 0.058(3) &  $-0.7(5)$  & 0.571 &    0.4\ \ \ \   \\
J1848+0351 & 0.066(2) &  $-2.1(4)$  & 0.665 &    4.1\ \ \ \   \\
J1850+0423 & 0.199(4) &  $-2.9(2)$  & 0.687 &    8.1\ \ \ \   \\
J1855+0205 & 0.193(5) &  $-1.4(3)$  & 1.313 &   24.5\ \ \ \   \\ [2pt]
J1856+0245 & 0.58(2) &  $-0.4(2)$  & 10.602 &   45.5\ \ \ \   \\
J1858+0346 & 0.190(9) &  $-3.0(4)$  & 0.494 &   10.2\ \ \ \   \\
J1904+0738 & 0.23(1) &  $-1.1(4)$  & 0.029 &    9.1\ \ \ \   \\
J1905+0902 & 0.097(6) &  $-1.7(5)$  & 0.263 &    8.5\ \ \ \   \\
J1909+0641 & 0.111(3) &  $-1.9(2)$  & 1.362 &    0.5\ \ \ \   \\ [2pt]
J1909+0749 & 0.226(8) &  $-1.3(3)$  & 0.098 &    1.0\ \ \ \   \\
J1916+1225 & 0.094(2) &  $-0.6(2)$  & 1.863 &    4.4\ \ \ \   \\
J1917+1737 & 0.047(1) &  $-1.4(4)$  & 0.004 &    2.3\ \ \ \   \\
J1919+1745 & 0.190(8) &  $-2.6(4)$  & 0.174 &    5.5\ \ \ \   \\
J1919+1314 & 0.224(6) &  $-1.3(3)$  & 2.378 &   53.9\ \ \ \   \\ [2pt]
J1924+1631 & 0.088(4) &  $-0.9(3)$  & 3.794 &   17.6\ \ \ \   \\
J1928+1746 & 0.278(8) &  \phs $0.0(3)$  & 1.866 &    9.5\ \ \ \   \\
J1934+2352 & 0.062(2) &  $-3.4(3)$  & 1.185 &    8.4\ \ \ \   \\
J1938+2010 & 0.103(9) &  $-1.9(7)$  & 0.233 &    8.8\ \ \ \   \\
J1940+2337 & 0.069(4) &  \phs $0.1(7)$  & 1.911 &    5.0\ \ \ \   \\ [2pt]
J1941+2525 & 0.240(8) &  $-0.8(3)$  & 0.097 &   22.6\ \ \ \   \\
J1946+2535 & 0.480(7) &  $-1.6(2)$  & 4.376 &   30.9\ \ \ \   \\
J1947+1957 & 0.081(6) &  $-1.5(8)$  & 0.027 &    3.4\ \ \ \   \\
J1948+2551 & 0.622(7) &  $-0.7(1)$  & 5.619 &   47.9\ \ \ \   \\
J1949+2306 & 0.097(4) &  $-3.3(3)$  & 1.089 &    7.8\ \ \ \   \\ [2pt]
J1953+2732 & 0.106(7) &  $-3.0(6)$  & 0.748 &    9.5\ \ \ \   \\
J2005+3547 & 0.252(9) &  $-2.2(4)$  & 0.082 &   \tablenotemark{a} \ \ \ \ \  \\
J2006+3102 & 0.27(1) &  $-1.8(4)$  & 0.157 &    6.4\ \ \ \   \\
J2007+3120 & 0.123(5) &  $-2.9(4)$  & 0.291 &    6.1\ \ \ \   \\
J2009+3326 & 0.15(1) &  $-4.1(6)$  & 0.261 &   19.7\ \ \ \   \\ [2pt]
J2010+2845 & 0.43(2) &  $-2.1(3)$  & 0.373 &    9.9\ \ \ \   \\
J2010+3230 & 0.118(5) &  $-3.5(3)$  & 0.902 &   31.6\ \ \ \   \\
J2011+3331 & 0.384(8) &  $-2.5(2)$  & 1.215 &   32.9\ \ \ \   \\
J2013+3058 & 0.067(5) &  $-3.0(7)$  & 0.253 &    2.4\ \ \ \   \\
J2018+3431 & 0.247(9) &  $-1.2(3)$  & 0.693 &   13.1\ \ \ \   \\
\enddata
\tablenotetext{a}{Unknown $L_{1400}$ due to unknown distance.}
\end{deluxetable}

\begin{deluxetable}{lr@{\hspace*{18pt}}rrrrcrrrr}
\tabletypesize{\scriptsize}
\tablecaption{\label{table:width}Pulse Widths
}
\tablehead{
\multicolumn{1}{c}{PSR\tablenotemark{a}} & \multicolumn{1}{c}{$P$ (ms)} &  \multicolumn{4}{c}{$W_{50}$ (ms)} & & \multicolumn{4}{c}{$W_{50}/P$} \\
\cline{3-6}\cline{8-11}
 & & 1170     & 1370     & 1470     & 1570     & & 1170     & 1370     & 1470     & 1570      \\
 & &      MHz &      MHz &      MHz &      MHz & &      MHz &      MHz &      MHz &      MHz  
}
\startdata
 0540+3207     &  524.3 &   11.0 &   11.7 &   12.0 &   12.4 & & 0.021 & 0.022 & 0.023 & 0.024 \\
 0628+0909     & 1241.4 &    7.6 &    7.7 &    8.0 &    8.4 & & 0.006 & 0.006 & 0.006 & 0.007 \\
 1848+0351     &  191.4 &    9.9 &    9.9 &   10.3 &   12.0 & & 0.052 & 0.052 & 0.054 & 0.063 \\
 1850+0423     &  290.7 &   24.6 &   24.3 &   24.7 &   24.6 & & 0.084 & 0.084 & 0.085 & 0.085 \\
 1855+0205     &  246.8 &   20.5 &    9.4 &    7.6 &    6.6 & & 0.083 & 0.038 & 0.031 & 0.027 \\[2pt]
 1856+0245     &   80.9 &   23.8 &   17.6 &   14.6 &   14.2 & & 0.295 & 0.218 & 0.180 & 0.175 \\
 1858+0346     &  256.8 &   41.4 &   33.9 &   36.7 &   32.3 & & 0.161 & 0.132 & 0.143 & 0.126 \\
 1904+0738     &  209.0 &    3.2 &    3.0 &    2.9 &    2.8 & & 0.015 & 0.014 & 0.014 & 0.013 \\
 1905+0902     &  218.3 &    3.9 &    3.7 &    3.9 &    3.9 & & 0.018 & 0.017 & 0.018 & 0.018 \\
 1909+0641     &  741.8 &    8.2 &    7.9 &    7.9 &    7.6 & & 0.011 & 0.011 & 0.011 & 0.010 \\[2pt]
 1909+0749M    &  237.2 &    5.6 &    4.7 &    5.2 &    5.0 & & 0.024 & 0.020 & 0.022 & 0.021 \\
 1909+0749I    &  237.2 &    4.0 &    3.7 &    3.7 &    3.4 & & 0.017 & 0.016 & 0.016 & 0.014 \\
 1916+1225     &  227.4 &    2.0 &    1.7 &    1.7 &    1.6 & & 0.009 & 0.008 & 0.007 & 0.007 \\
 1917+1737     &  334.7 &    6.1 &    5.5 &    5.9 &    5.7 & & 0.018 & 0.016 & 0.018 & 0.017 \\
 1919+1314     &  571.4 &   15.6 &   13.2 &   11.5 &   11.6 & & 0.027 & 0.023 & 0.020 & 0.020 \\[2pt]
 1919+1745     & 2081.3 &   63.6 &   62.4 &   62.4 &   62.5 & & 0.031 & 0.030 & 0.030 & 0.030 \\
 1924+1631     & 2935.2 &   35.1 &   27.1 &   27.2 &   24.1 & & 0.012 & 0.009 & 0.009 & 0.008 \\
 1928+1746     &   68.7 &    3.4 &    3.5 &    3.4 &    3.4 & & 0.050 & 0.051 & 0.050 & 0.049 \\
 1934+2352     &  178.4 &    4.5 &    3.7 &    4.2 &    3.3 & & 0.025 & 0.021 & 0.024 & 0.018 \\
 1938+2010     &  687.1 &   24.7 &   24.5 &   25.0 &   24.7 & & 0.036 & 0.036 & 0.036 & 0.036 \\[2pt]
 1940+2337     &  546.8 &   10.6 &   10.4 &    9.4 &   10.3 & & 0.019 & 0.019 & 0.017 & 0.019 \\
 1941+2525     & 2306.2 &   33.5 &   36.6 &   35.6 &   33.4 & & 0.015 & 0.016 & 0.015 & 0.015 \\
 1946+2535     &  515.2 &    9.5 &    8.7 &    8.6 &    8.5 & & 0.018 & 0.017 & 0.017 & 0.016 \\
 1947+1957     &  157.5 &    6.9 &    6.1 &    6.5 &    5.9 & & 0.044 & 0.039 & 0.041 & 0.038 \\
 1948+2551     &  196.6 &    9.1 &    9.0 &    9.2 &    9.0 & & 0.046 & 0.046 & 0.047 & 0.046 \\[2pt]
 1949+2306     & 1319.4 &   17.8 &   17.8 &   19.6 &   21.5 & & 0.013 & 0.014 & 0.015 & 0.016 \\
 1953+2732     & 1334.0 &   29.7 &   29.0 &   26.3 &   26.7 & & 0.022 & 0.022 & 0.020 & 0.020 \\
 2005+3547     &  615.0 &   23.3 &   19.6 &   20.8 &   18.9 & & 0.038 & 0.032 & 0.034 & 0.031 \\
 2006+3102     &  163.7 &   12.2 &   11.8 &   11.8 &   11.8 & & 0.074 & 0.072 & 0.072 & 0.072 \\
 2007+3120     &  608.2 &    8.7 &    8.9 &    8.7 &    8.8 & & 0.014 & 0.015 & 0.014 & 0.015 \\[2pt]
 2009+3326     & 1438.4 &   24.5 &   25.8 &   26.7 &   29.7 & & 0.017 & 0.018 & 0.019 & 0.021 \\
 2010+2845     &  565.4 &   14.6 &   14.1 &   13.6 &   13.5 & & 0.026 & 0.025 & 0.024 & 0.024 \\
 2010+3230     & 1442.4 &   30.6 &   28.1 &   28.8 &   26.8 & & 0.021 & 0.019 & 0.020 & 0.019 \\
 2011+3331     &  931.7 &   30.7 &   30.6 &   30.6 &   30.1 & & 0.033 & 0.033 & 0.033 & 0.032 \\
 2013+3058     &  276.0 &    3.0 &    2.8 &    2.9 &    3.0 & & 0.011 & 0.010 & 0.010 & 0.011 \\[2pt]
 2018+3431     &  387.7 &    6.8 &    6.6 &    6.7 &    6.5 & & 0.017 & 0.017 & 0.017 & 0.017 \\
\enddata
\tablenotetext{a}{Main pulse and interpulse denoted by M and I, respectively.}
\end{deluxetable}

\addtocounter{figure}{-1}

\begin{figure*}[p]
\begin{center}
\includegraphics[width=6.0in]{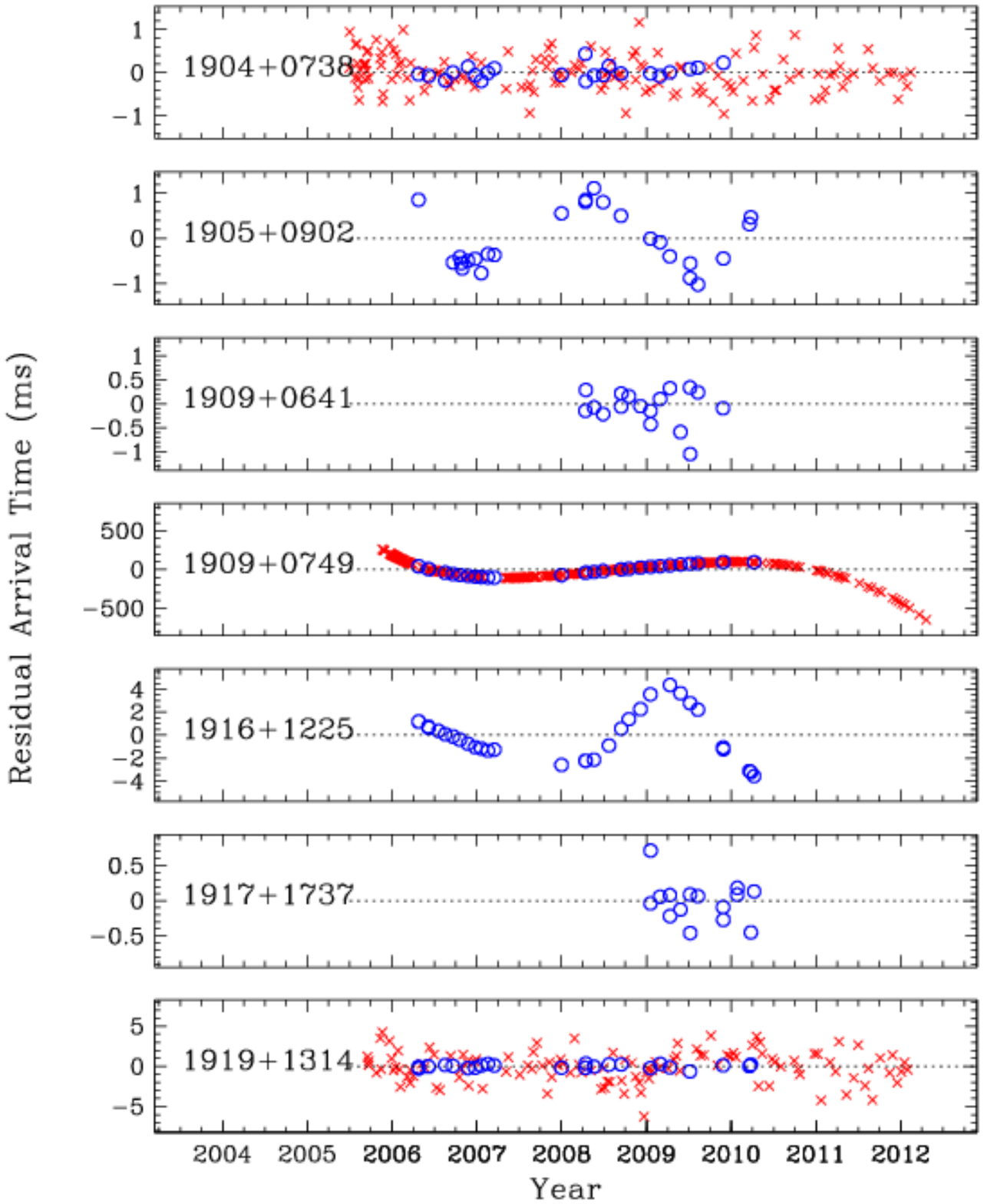}
\caption{Part (b).}
\end{center}
\end{figure*}

\addtocounter{figure}{-1}

\begin{figure*}[p]
\begin{center}
\includegraphics[width=6.0in]{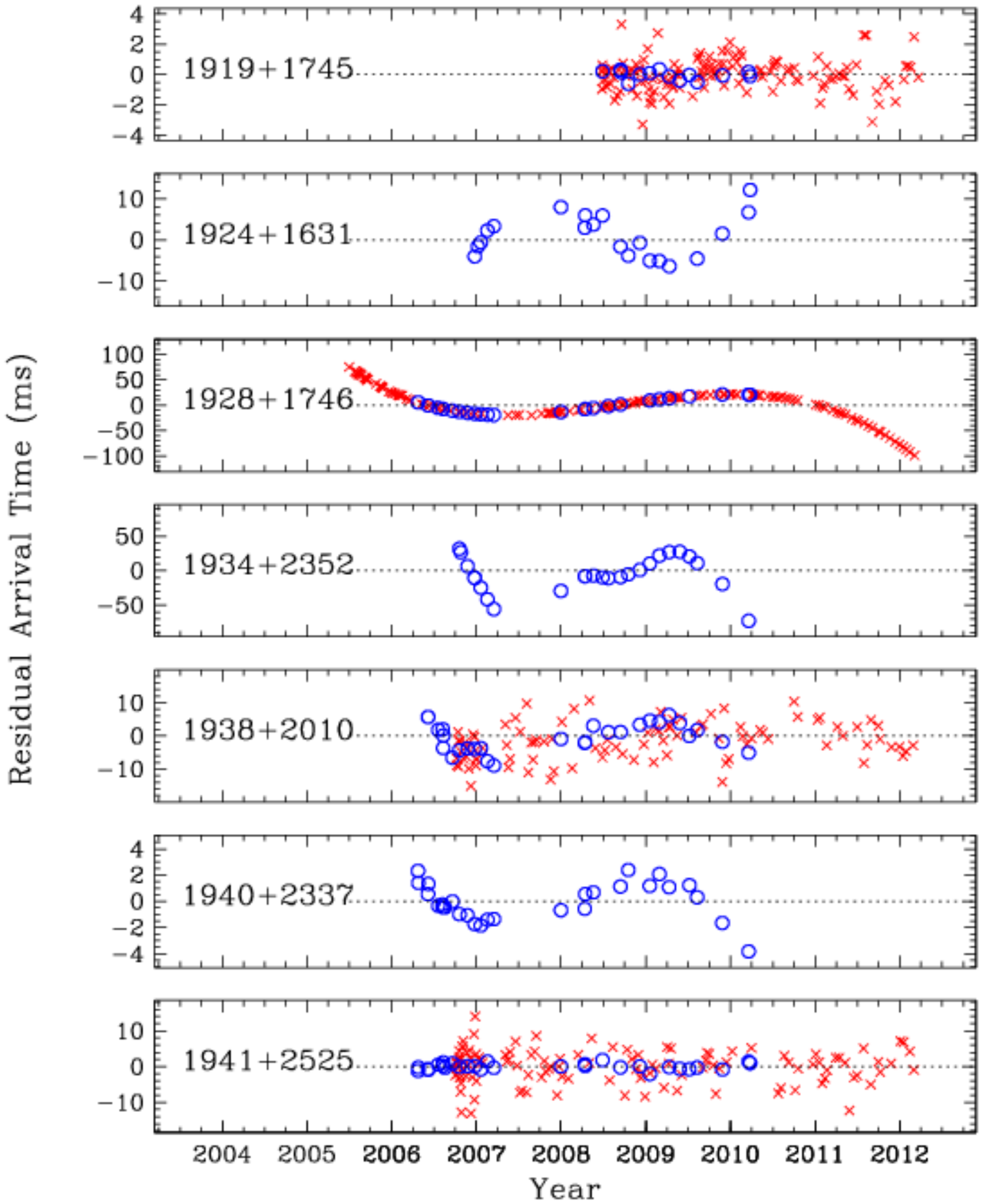}
\caption{Part (c).}
\end{center}
\end{figure*}

\addtocounter{figure}{-1}

\begin{figure*}[p]
\begin{center}
\includegraphics[width=6.0in]{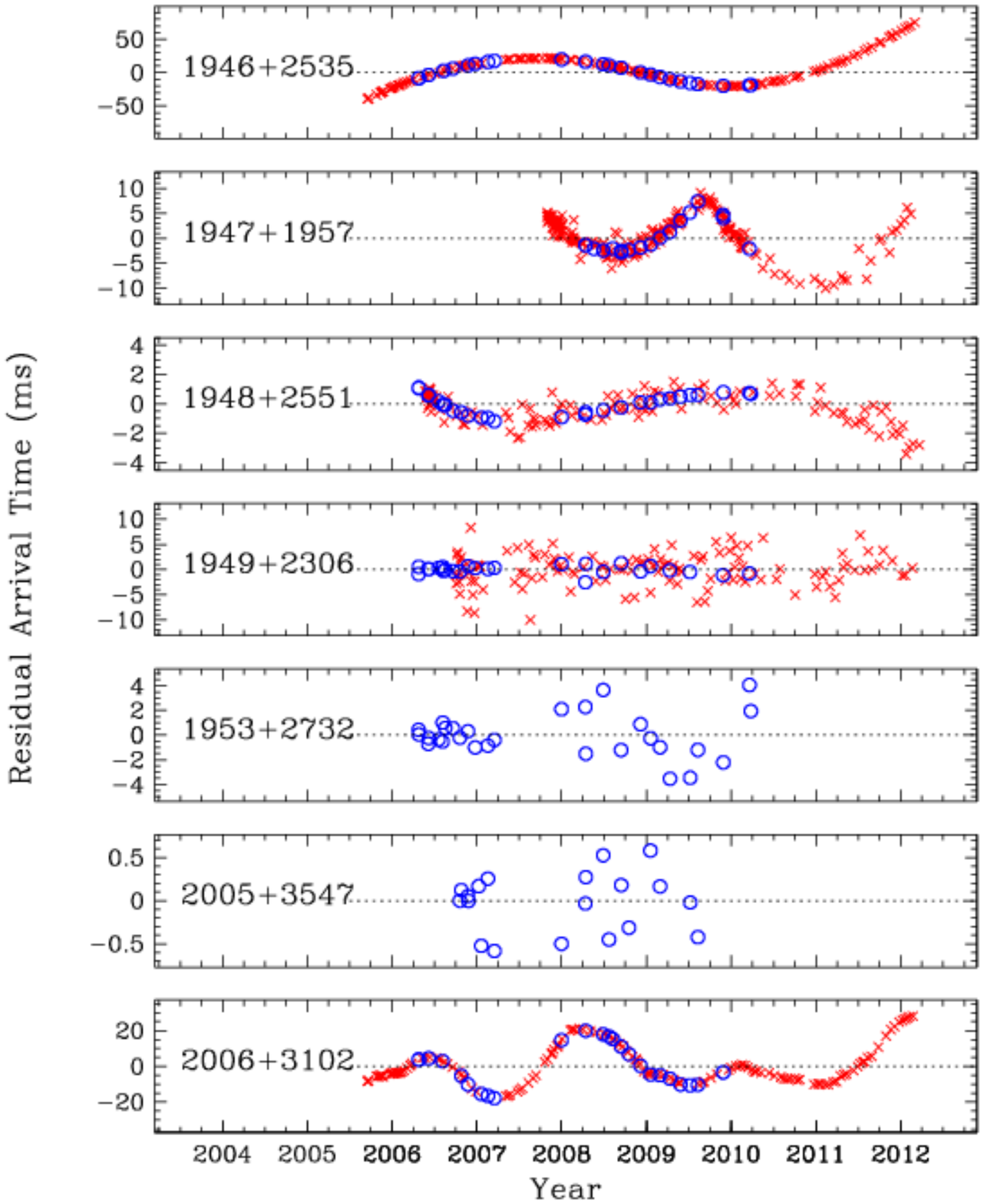}
\caption{Part (d).}
\end{center}
\end{figure*}

\addtocounter{figure}{-1}

\begin{figure*}[p]
\begin{center}
\includegraphics[width=6.0in]{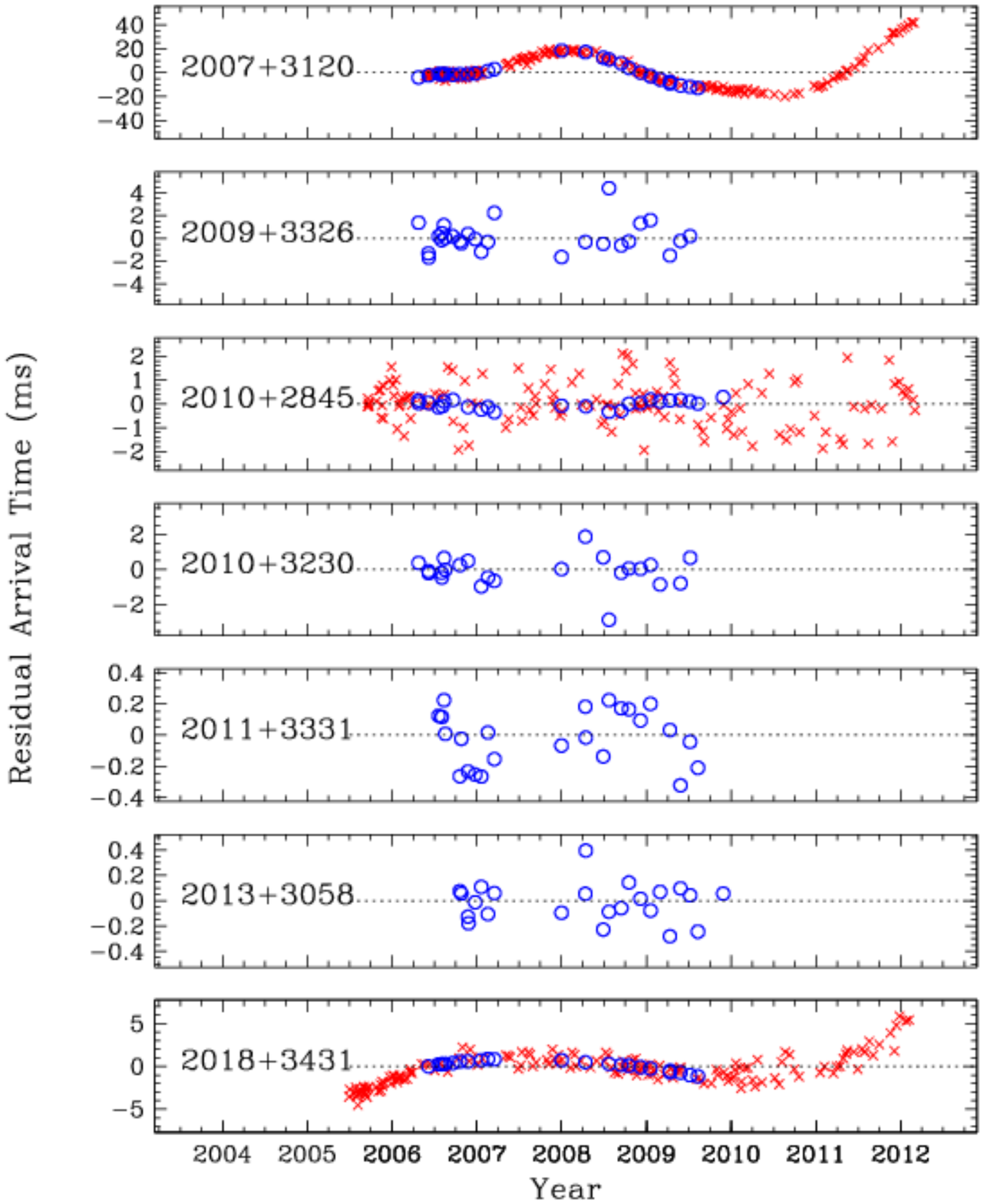}
\caption{Part (e).}
\end{center}
\end{figure*}

\begin{figure*}[p]
\begin{center}
\includegraphics[width=0.80\linewidth]{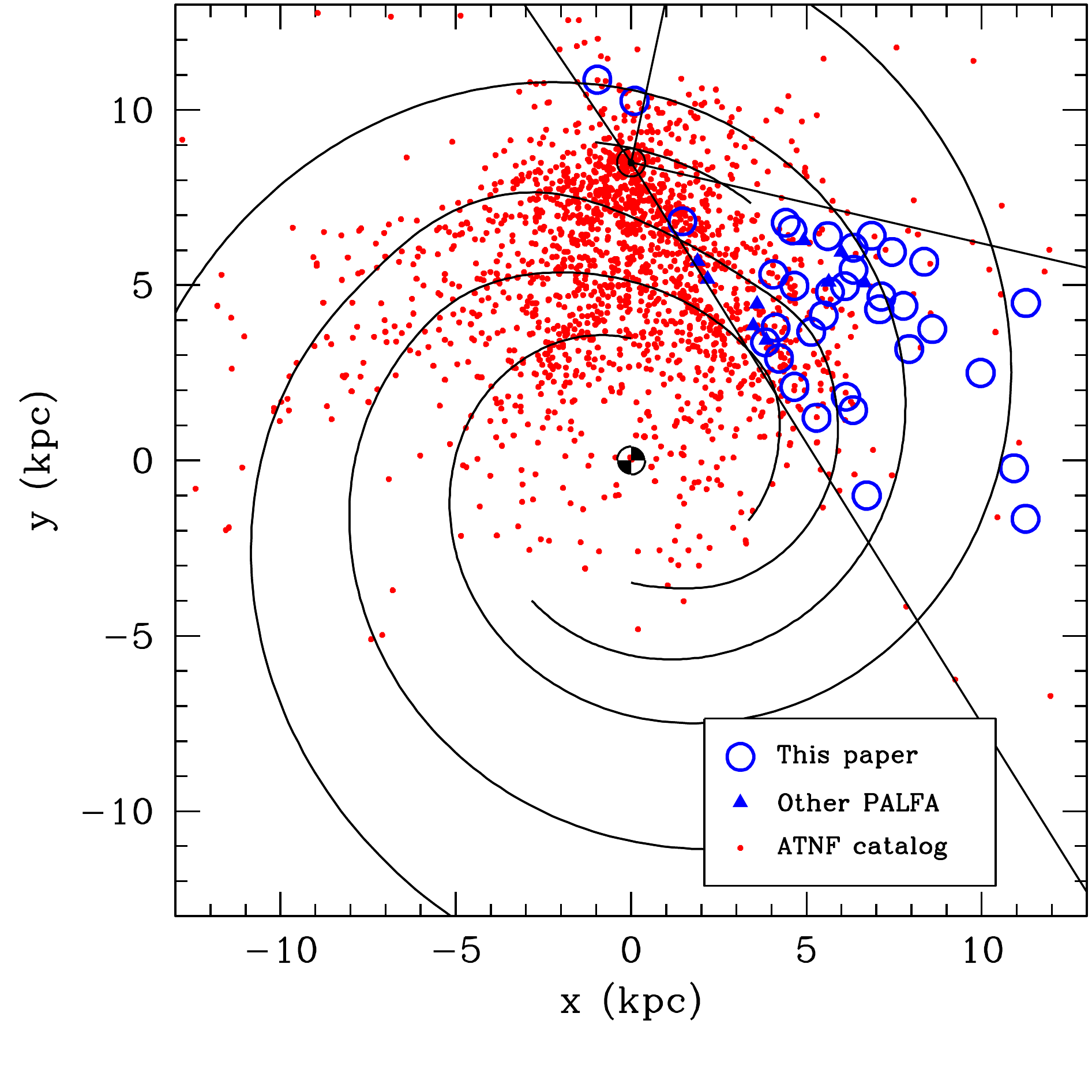}
\caption{\label{fig:galplot}Pulsar positions projected onto
the Galactic plane.  Pulsar positions are based on the coordinates and
distances given in Table~\ref{table:scatter}.  
The position of the Sun is indicated by a small circle and
dot at $(x,y)=(0,8.5\,{\rm kpc})$.
Straight lines emerging from the Sun indicate regions of low Galactic
latitude within the declination range of the Arecibo telescope.
Curved lines indicate Galactic spiral arms according to the NE2001
electron density model \citep{cl02}).  
Blue circles 
indicate pulsars in the present paper; 
blue triangles 
indicate other published pulsars discovered in the PALFA survey; and 
red dots 
are all known pulsars in the Galactic disk
\citep{hm13}.
}
\end{center}
\end{figure*}

\begin{figure*}[p]
\begin{center}
\includegraphics[width=1.00\linewidth]{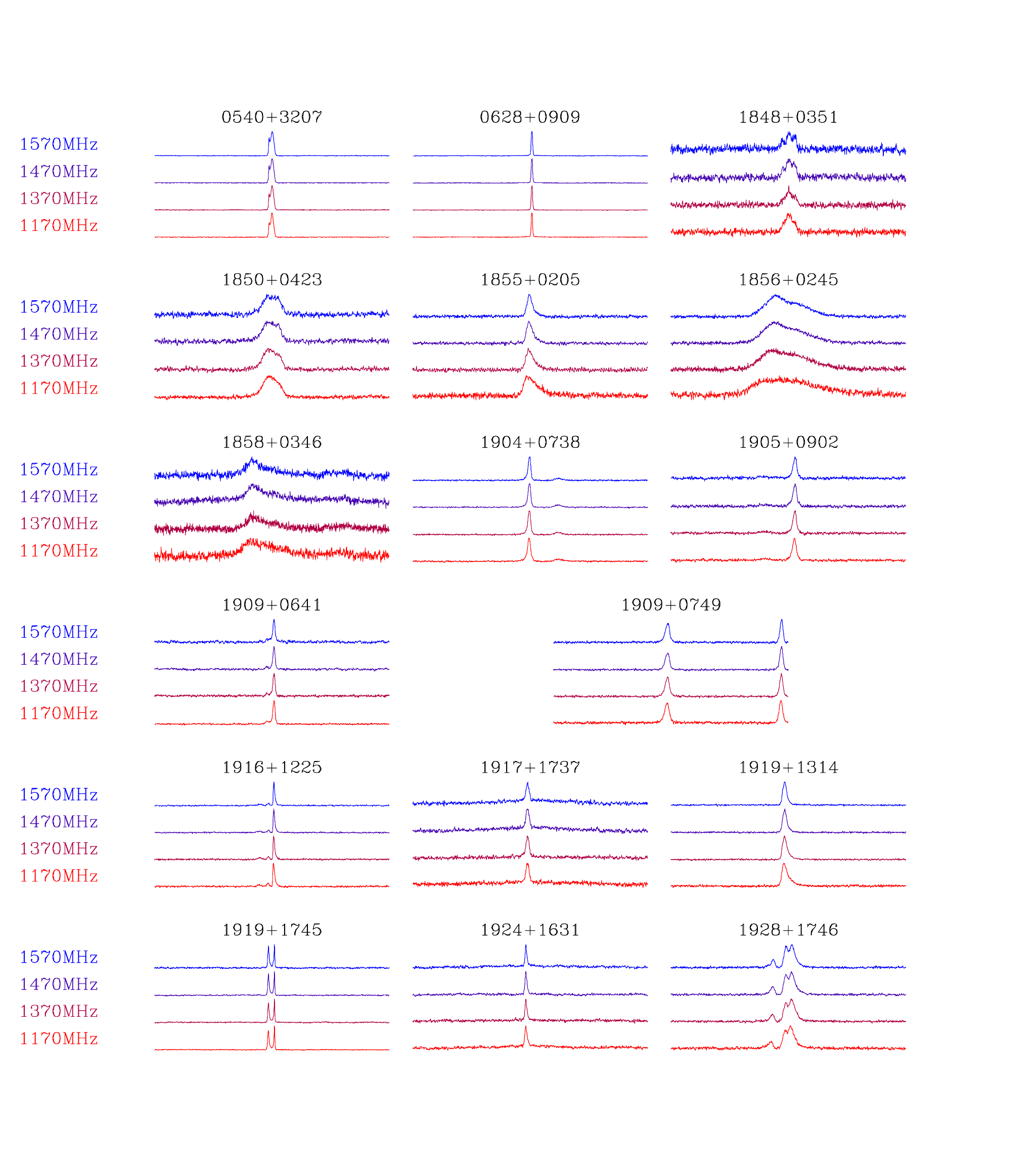}
\caption{\label{fig:profilea}%
Part (a).
Profiles of the pulsars in each of the four observational subbands are
given in (a) and (b).  Each profile
is a full pulse period.  Profiles are sums of all Arecibo data records
which produced valid TOAs for a given pulsar at a given frequency.  Data
records were aligned using the final timing solution for each pulsar.
Plots are total intensity (summed polarizations).  See
Figures~\ref{fig:profilecloseupa}(a) and (b) for 
expanded views of the pulse peaks.
}
\end{center}
\end{figure*}

\addtocounter{figure}{-1}

\begin{figure*}[p]
\begin{center}
\includegraphics[width=1.00\linewidth]{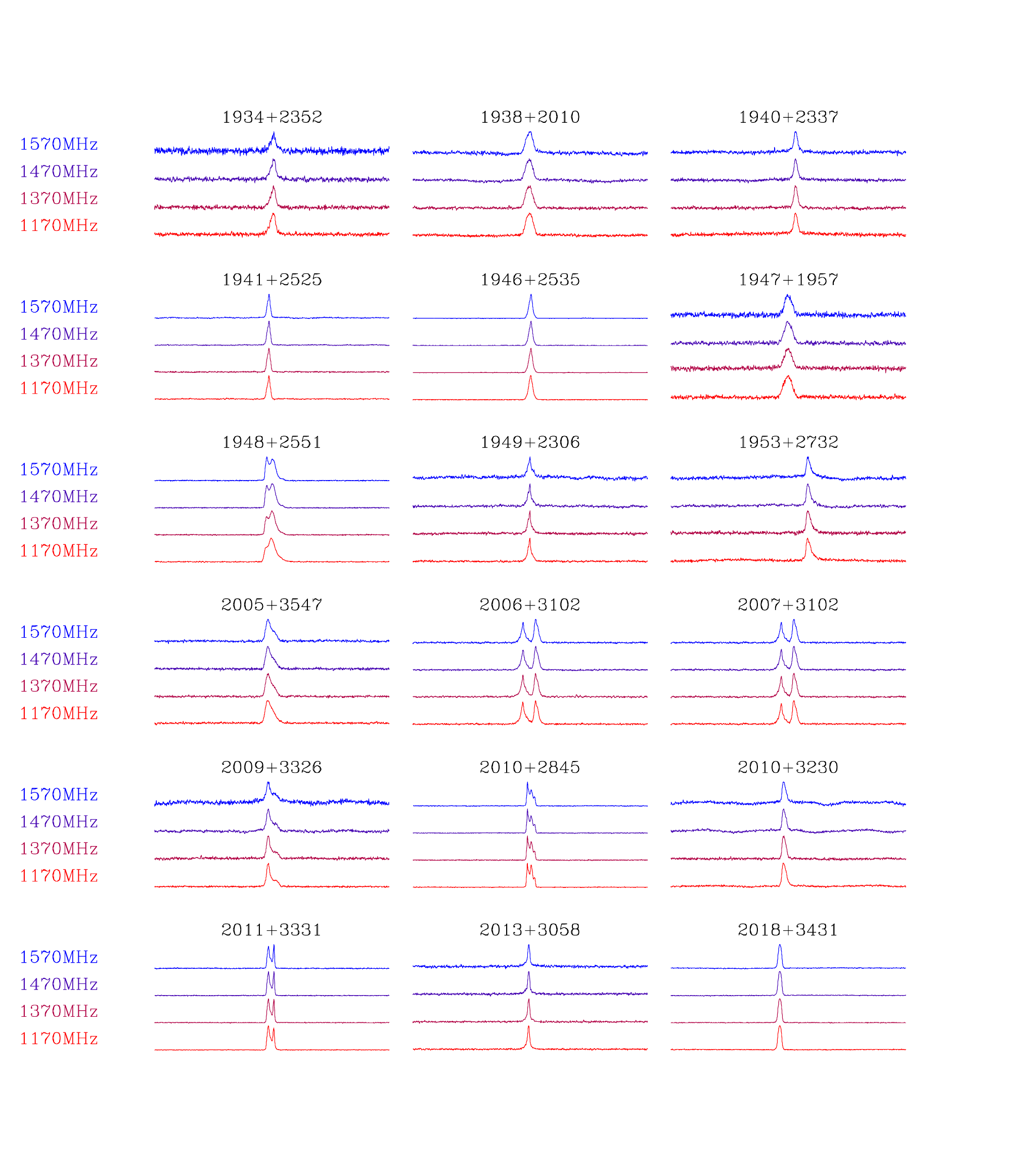}
\caption{\label{fig:profileb}%
Part (b).
}
\end{center}
\end{figure*}

\begin{figure*}[p]
\begin{center}
\includegraphics[width=1.00\linewidth]{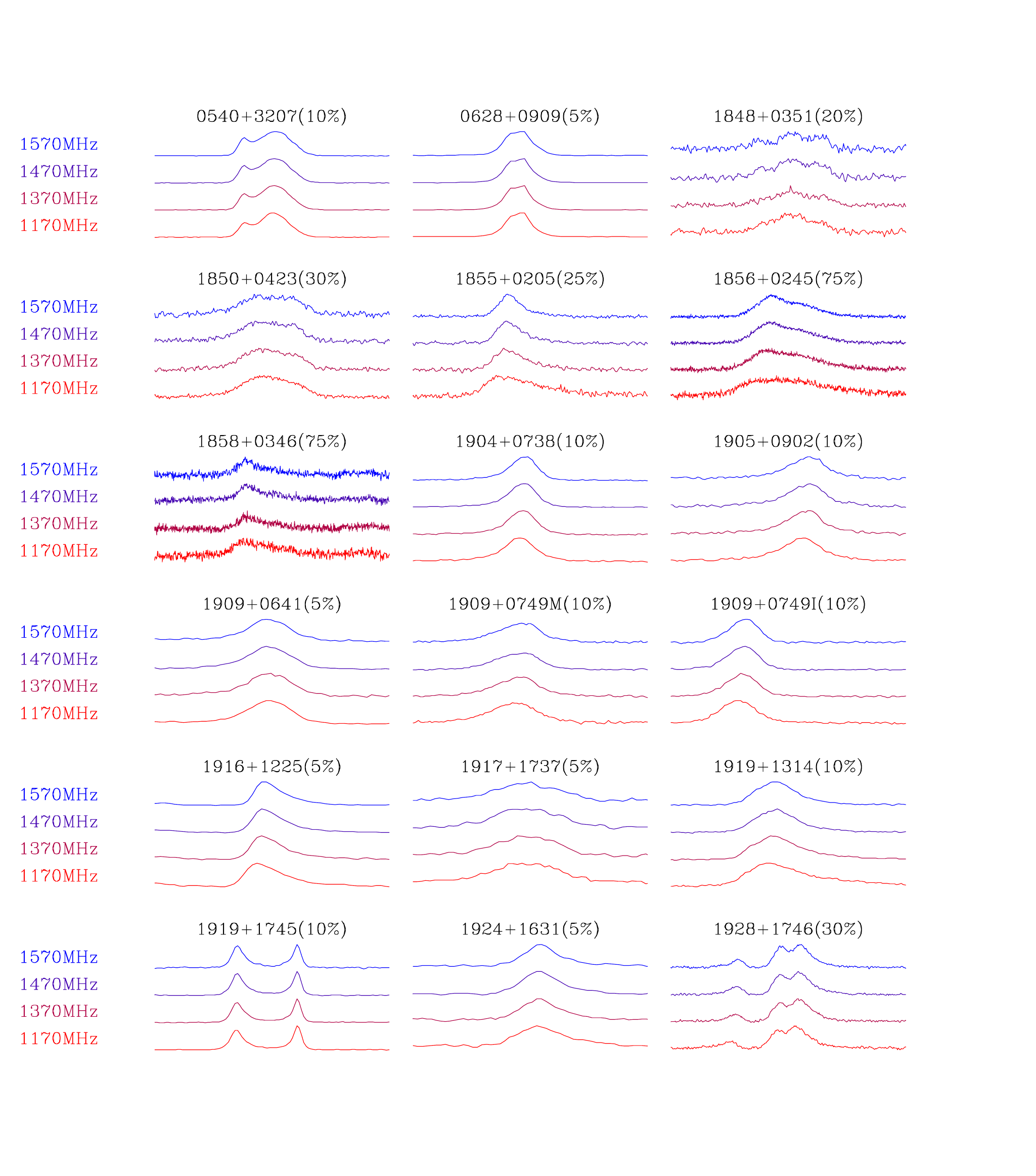}
\caption{\label{fig:profilecloseupa}%
Part (a).
Close-up of the pulse peak of each pulse profile in
Figures~\ref{fig:profilea}(a) and (b).  The horizontal range varies from pulsar to pulsar
depending on the pulse width and is given above each set of profiles, e.g.,
``0540+3207(10\%)'' means that the horizontal extent of the profiles of
0540+3207 is 10\% of the pulse period.  Two plots are shown for 1909+0749,
labeled by ``M'' and ``I,'' for main pulse and interpulse; these correspond
to the leading (``M'') and trailing components (``I'') of the profile shown
in Figure~\ref{fig:profilea}(a).  These two components are similar in total
energy, and the designation of ``M'' and ``I'' is arbitrary.  They are
plotted on the same vertical scale, and the centers of the two plots are
separated by half a pulse period.
}
\end{center}
\end{figure*}

\addtocounter{figure}{-1}

\begin{figure*}[p]
\begin{center}
\includegraphics[width=1.00\linewidth]{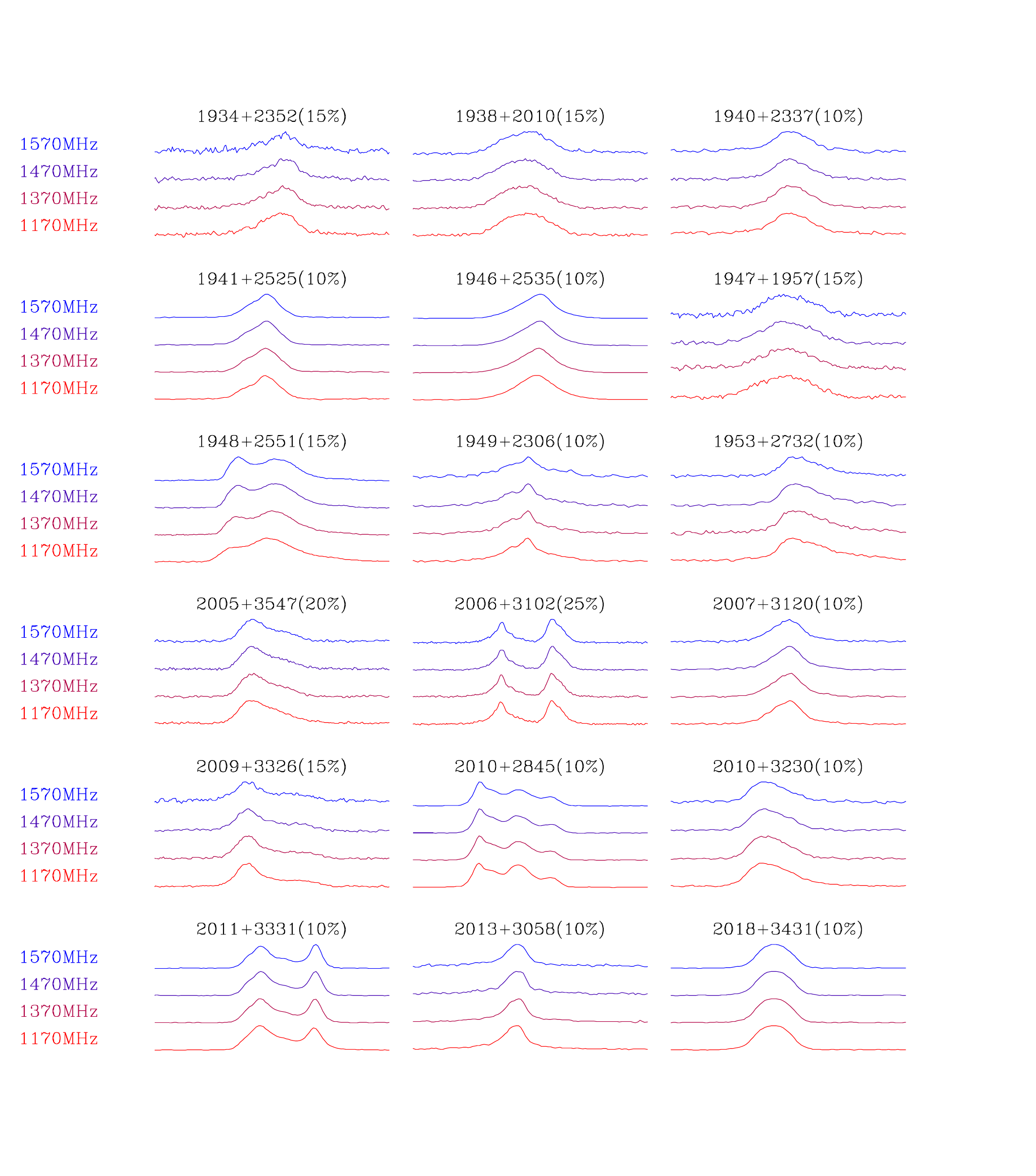}
\caption{\label{fig:profilecloseupb}%
Part(b).
}
\end{center}
\end{figure*}

\begin{figure*}[p]
\begin{center}
\includegraphics[trim=50 40 50 20, clip,scale=0.45]{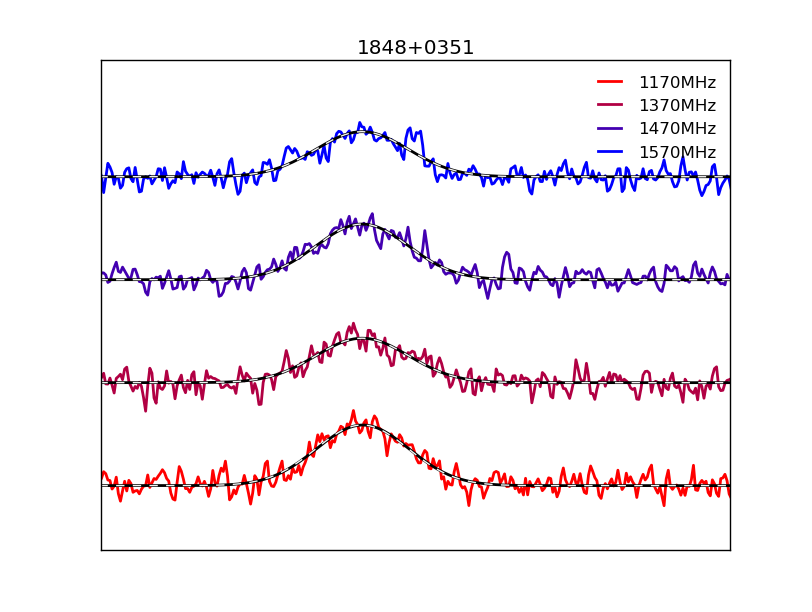}
\includegraphics[trim=50 40 50 20, clip,scale=0.45]{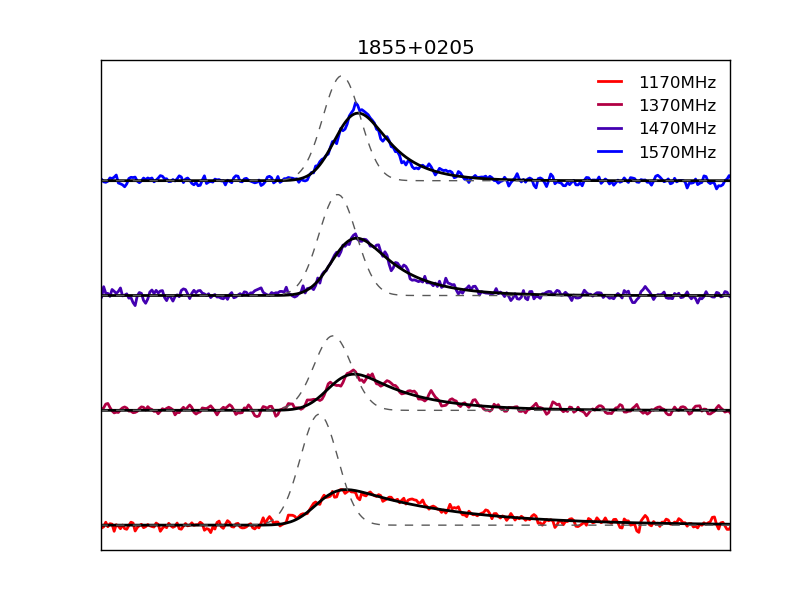}
\includegraphics[trim=50 40 50 20, clip,scale=0.45]{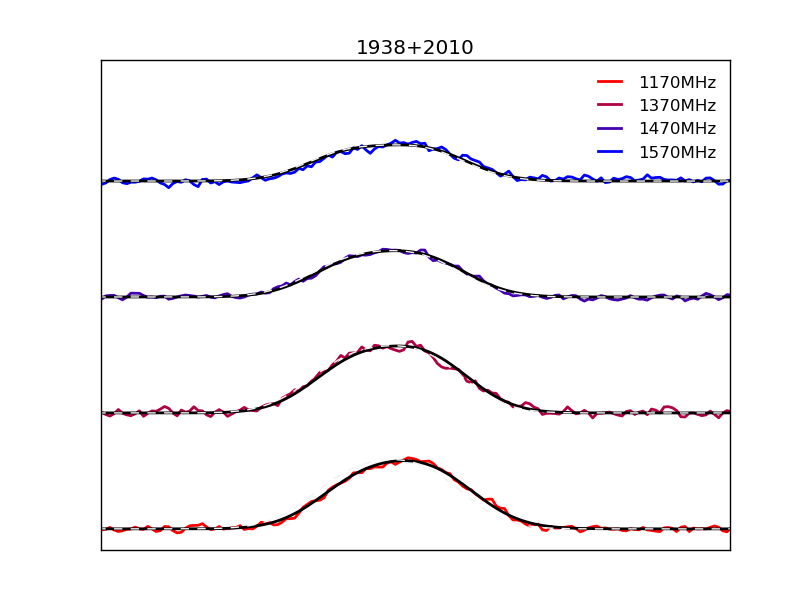}
\includegraphics[trim=50 40 50 20, clip,scale=0.45]{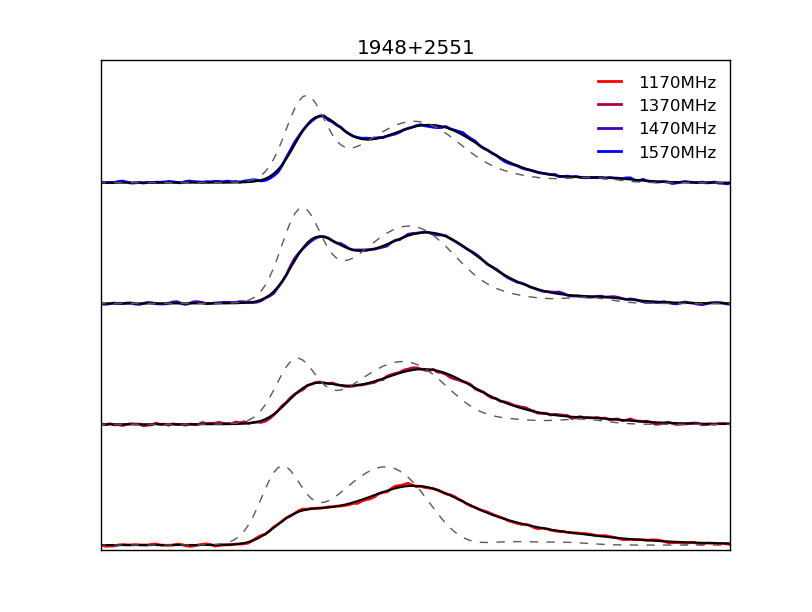}
\includegraphics[trim=50 40 50 20, clip,scale=0.45]{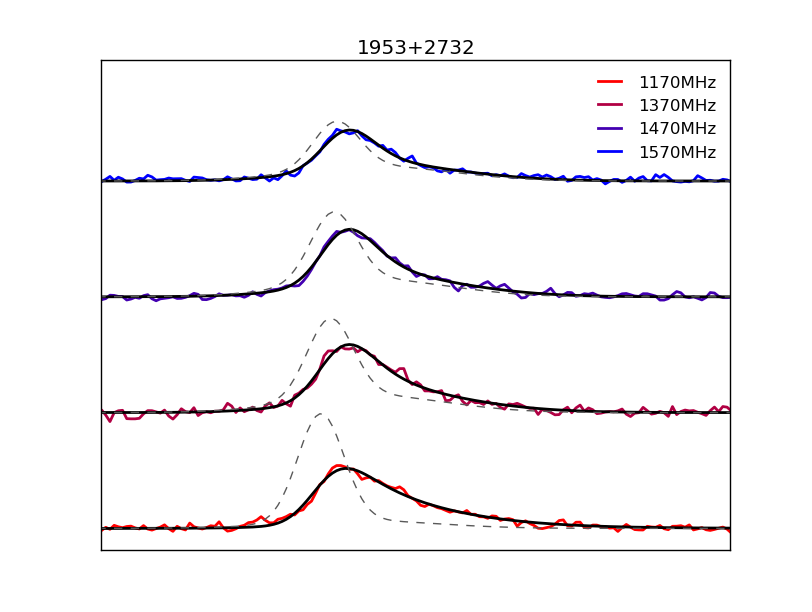}
\includegraphics[trim=50 40 50 20, clip,scale=0.45]{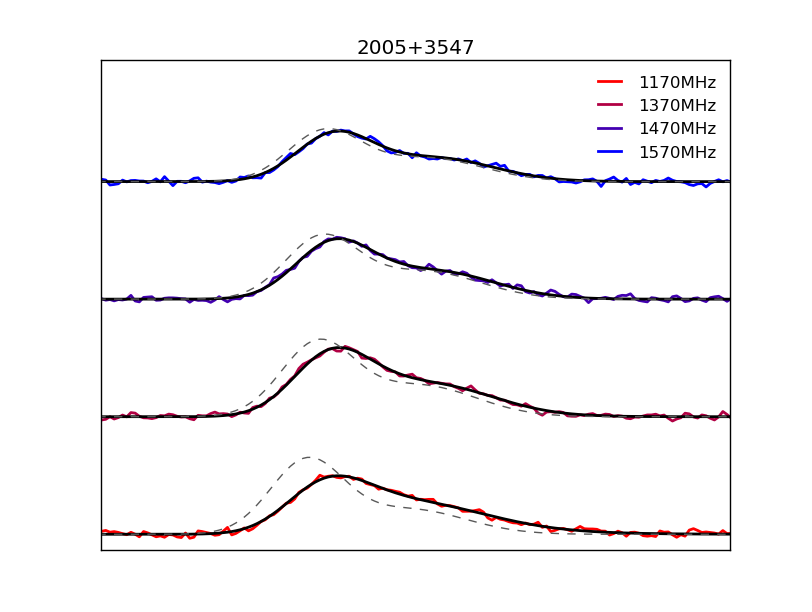}
\end{center}
\caption{\label{fig:scatter} 
Sample pulsar scattering model fits.  Top: PSR~J1848+0351, PSR~J1855+0205;
middle:  PSR~J1938+2010, PSR~J1948+2551; bottom: PSR~J1953+2732,
PSR~J2005+3557.  For each pulsar, closeups of the pulse peak are shown for
the four observing subbands (top to bottom: 1570, 1470, 1370, and 1170
MHz).  Superimposed on each profile are the profile models for that
frequency both with scattering (black line) and without scattering (gray
line).  Of the pulsars shown here, J1855+0205 and J1953+2732 show strong
scattering; J1848+0351 and J1938+2010 show almost no scattering; and
J1948+2551 and J2005+3547 have high covariance between the Gaussian profile
model and the scattering tail.
}
\end{figure*}

\begin{figure*}[p]
\begin{center}
\includegraphics[width=0.80\linewidth]{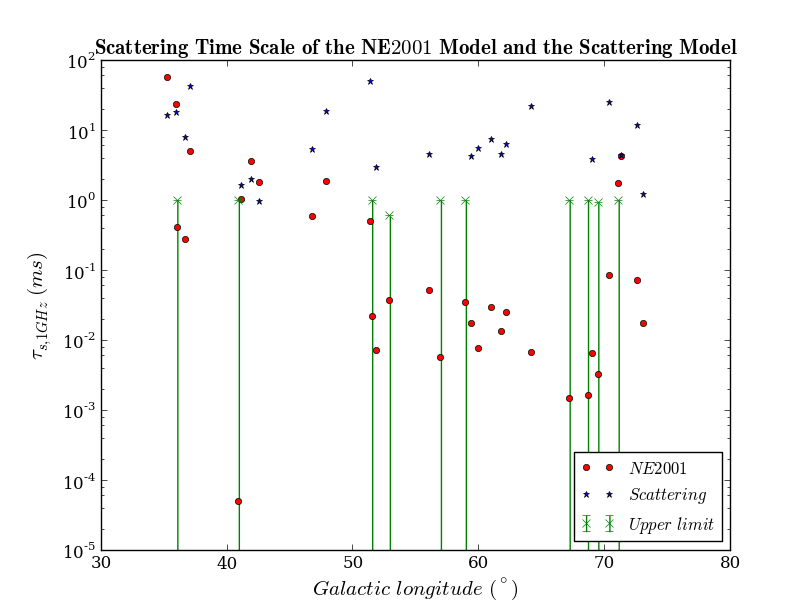}
\end{center}
\caption{\label{fig:scatter_gal}
Scattering time scales of the pulsars, comparing our measured time 
scales with the predictions of the NE2001 electron
density model \citep{cl02}.
}
\end{figure*}

\clearpage

\begin{figure*}[p!]
\begin{center}
\includegraphics[width=0.42\linewidth]{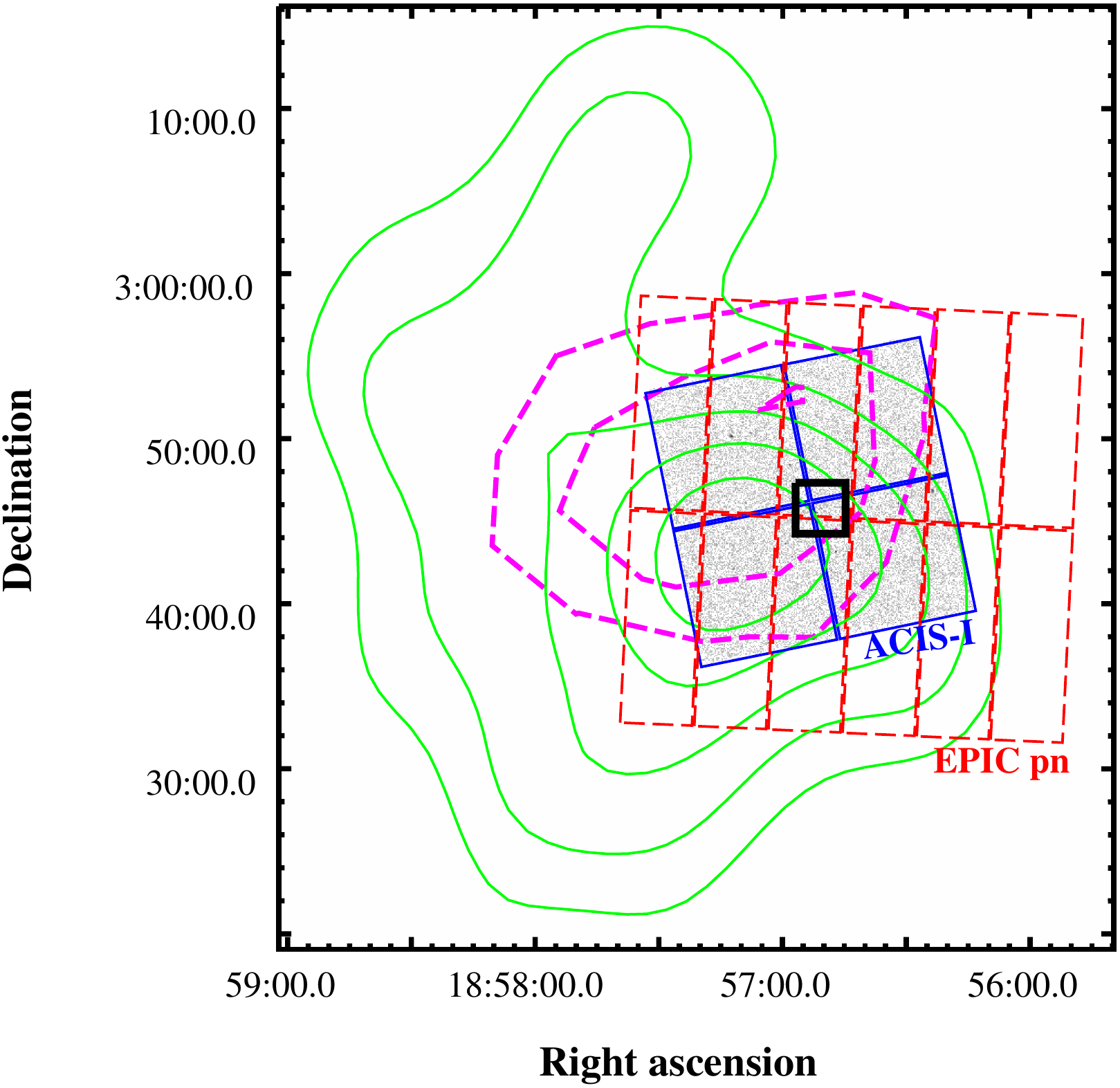}
\hspace*{0.05\linewidth}
\includegraphics[width=0.48\linewidth]{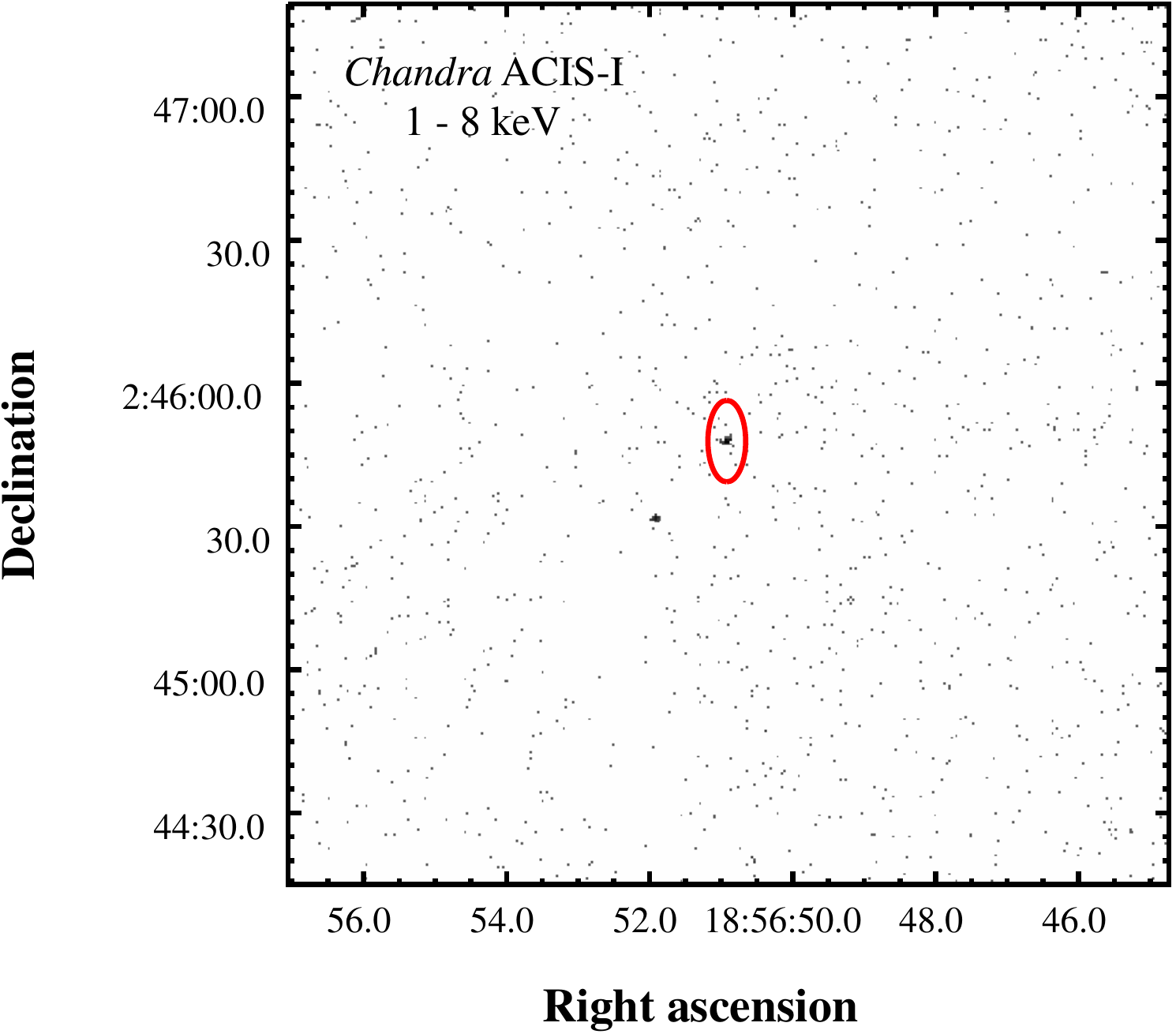}
\end{center}

\caption{\label{fig:1856chandra}
{\it Left.}  
Solid green contours:
HESS 1857+026 TeV source.  
Dashed magenta contours:
\textit{Fermi} Large Area Telescope source coincident with HESS J1857+026
\citep{rgv+12}.  Contours correspond to test statistic source significance
levels of 25, 16, and 9 (5, 4, and 3 $\sigma$).  
Dashed red rectangle:
{\it
XMM-Newton} EPIC pn field of view.  
Dashed blue square:
{\it Chandra} ACIS-I
1--8 keV image.  Black square: region shown in right image; pulsar position
is in the center.  {\it Right.}  {\it Chandra} ACIS-I image in the 1-8 keV
band centered on PSR~J1856+0245.  The 
red ellipse 
shows the position
uncertainty from the pulsar timing solution (Table~\ref{table:parameters}).
Pixel randomization was removed from the pipeline processing.  
}
\end{figure*}

\end{document}